\documentclass[11pt, oneside]{article}\usepackage[]{graphicx}\usepackage[]{xcolor}
\makeatletter
\def\maxwidth{ %
  \ifdim\Gin@nat@width>\linewidth
    \linewidth
  \else
    \Gin@nat@width
  \fi
}
\makeatother

\definecolor{fgcolor}{rgb}{0.345, 0.345, 0.345}

\usepackage{framed}
\makeatletter
 {\par\unskip\endMakeFramed%
 \at@end@of@kframe}
\makeatother

\definecolor{shadecolor}{rgb}{.97, .97, .97}
\definecolor{messagecolor}{rgb}{0, 0, 0}
\definecolor{warningcolor}{rgb}{1, 0, 1}
\definecolor{errorcolor}{rgb}{1, 0, 0}
\newenvironment{knitrout}{}{} 

\usepackage{alltt}
\usepackage{geometry}
\usepackage{graphicx}
\usepackage{amssymb}
\usepackage{amsmath}
\usepackage{url}
\usepackage{adjustbox}
\usepackage{comment}
\usepackage{subfigure}
\usepackage{booktabs}
\usepackage{authblk}
\usepackage{setspace}
\usepackage{breakcites}
\usepackage{float}
\usepackage{xspace}
\usepackage[authoryear, round]{natbib}
\usepackage{subfig}
\usepackage{booktabs}
\usepackage{lineno}

\usepackage{xcolor}

\makeatletter

\title{An algorithm for forensic toolmark comparisons}

 \author[1,2,4]{Maria Cuellar}
 
 \author[2]{Sheng Gao}
 
 \author[3,4]{Heike Hofmann}
 
 \affil[1]{Department of Criminology, University of Pennsylvania, 3718 Locust Walk, Philadelphia, PA, 19104, United States}
 
 \affil[2]{Department of Statistics and Data Science, Wharton School, University of Pennsylvania, Walnut Street, Philadelphia, PA 19104, United States}
 
 \affil[3]{Department of Statistics, Iowa State University, Snedecor Hall, 2438 Osborn Drive, Ames, IA, 50011, United States}
 
 \affil[4]{Center for Statistics and Applications in Forensics Evidence (CSAFE), Iowa State University, 613 Morrill Road, Ames, IA, 50011, United States}

\IfFileExists{upquote.sty}{\usepackage{upquote}}{}
\begin{document}

\maketitle

\begin{abstract}


Forensic toolmark analysis traditionally relies on subjective human judgment, leading to inconsistencies and lack of transparency. The multitude of variables, including angles and directions of mark generation, further complicates comparisons. To address this, we first generate a dataset of 3D toolmarks from various angles and directions using consecutively manufactured slotted screwdrivers. By using PAM clustering, we find that there is clustering by tool rather than angle or direction. Using Known Match and Known Non-Match densities, we establish thresholds for classification. Fitting Beta distributions to the densities, we allow for the derivation of likelihood ratios for new toolmark pairs. With a cross-validated sensitivity of 98\% and specificity of 96\%, our approach enhances the reliability of toolmark analysis. This approach is applicable to slotted screwdrivers, and for screwdrivers that are made with a similar production method. With data collection of other tools and factors, it could be applied to compare toolmarks of other types. This empirically trained, open-source solution offers forensic examiners a standardized means to objectively compare toolmarks, potentially decreasing the number of miscarriages of justice in the legal system.
\end{abstract}

\section{Introduction}

Tools like screwdrivers, crowbars, or wire cutters are often used during the commission of a crime, such as breaking into a property or making a explosive device. The goal of forensic toolmark examiners is to determine whether a suspected tool, if available, made the mark. An examiner might also compare two marks of unknown source to determine whether they came from the same source \citep{baldwinbook}. The examiner generates test marks with the suspected tool at different angles and directions in a laboratory, and then compares the crime scene mark and test marks \citep{petraco}. The result of this conclusion can then be used as evidence in a legal case \citep{nichols1, nichols2}.

The decision about whether two toolmarks were made by the same tool relies on subjective, human judgment. Toolmark examiners compare the marks subjectively by using a comparison light microscope, which depicts the striation marks as light and dark patterns in 2D \citep{petraco}. Then, the examiner must decide whether the marks were made by the same source or different source by determining whether the ``surface contours of two toolmarks are in `sufficient agreement''' based on the examiner's opinion that another tool could not have made the marks \citep{afte}. Subjective methods are susceptible to ``human error, bias, and performance variability across examiners''~\citep{pcast}, and these errors have contributed to wrongful convictions as well as miscarriages of justice. Out of the 3,290 exonerations recorded in the National Registry of Exonerations as of March 2023, in 24\% of them forensic science was a contributing factor to the wrongful conviction~\citep{nationalregistry}. 

For this reason, researchers~\citep{kafadar2019} and government reports~\citep{nas2009, pcast} have recommended that objective methods be used for forensic comparisons since these tend to yield greater accuracy and consistency, and since ``a process that has been defined with quantifiable, objective steps is easier to validate"~\citep{kafadar2019}. Furthermore, 2D comparisons made with a light microscope are sensitive to the lighting parameters, and they do not have precise information about the depth of the striations~\cite{vorburger-light}. Because 3D data contains precise information about the striation depth, it is likely to yield more accurate results in comparison with 2D data~\citep{vorburger, baiker2014}.

Although some~\citep{afte} have cited research on firearm 3D algorithms (see \citet{hare, chu, vorburger2011, tai2018fully}, to name a few) to argue that there is extensive research about non-firearm toolmarks, the research on firearms is not directly relevant to non-firearm toolmarks. This is because of the nature of the data, and in fact, non-firearm toolmarks are more difficult to analyze than firearm marks. In firearms the shape of a toolmark does not depend on the firearm user's decisions. There is only one way, without much variabilty, to shoot the firearm. In contrast, in non-firearm toolmarks, factors such as the angle of attack and the direction in which the mark was made affect the shape of the toolmark~\citep{petraco, baldwinbook}. This is sometimes called the ``degrees of freedom" problem of non-firearm toolmarks, and it adds difficulty to the comparison. If marks made by a single tool at the same conditions vary drastically from one replicate to the next, specifically if they are more different from each other than the marks made by two different (e.g., consecutively manufactured) tools, then toolmark comparisons will not be successful in general. Research on objective methods in non-firearm toolmarks includes the complexity of the degrees of freedom problem \citep{baiker2015, lockmorris2013, macziewski, spotts, spottsangle}. For instance, \cite{baiker2015} study how much marks change as the angle of attack of the tool changes. 

In this article, we present an open-source algorithm, an objective method for comparison of 3D scans of striated toolmarks. We made three contributions to the field. First, we generated three databases of toolmarks from consecutively manufactured flat-head unused screwdrivers: one to study the variability within and between tools at a fixed angle/direction, one to study the variability within and between angles of attack (80, 70, 60 with respect to the surface), and one to study the variability within and between directions of tool travel (pushing and pulling). We use the GelSight portable handheld 3D scanner, which  measures the 3D topography of a solid surface using elastomeric tactile sensor technology. This dataset is available to researchers.

Second, we ran a PAM clustering algorithm \cite{kaufman1990partitioning} on the three databases to determine whether the similarity within source (tool-side) was higher than the similarity between sources, at a fixed angle/direction, when varying angle, and when varying direction. Then we used what we learned from the clustering step to generate the Known-Match and Known-Non-Match densities. We fit a Beta parametric distribution to these, and used an ensemble method \citep{fedesproject}, to generate likelihood ratios. Thus, a new pair or marks can be compared, and a likelihood ratio can be generated, using our method. We provide R software, an open-source implementation of this process, to make it openly and freely accessible to forensic examiners and researchers.

Third, using our method, we find that very short signals (under 1.5 mm long) cannot be compared reliably.

Section \ref{sec:previouswork} describes previous work on this topic, Section \ref{sec:data} describes the data generation process, Section \ref{sec:methods} describes the methods used, Section \ref{sec:results} describes our results, Section \ref{ref:performance} describes the method's performance, and Section \ref{sec:discussion} gives a discussion. Regarding terminology, we use the term {\it tool} to mean a single screwdriver, {\it source} to mean the side of a screwdriver tip (side A or B), {\it mark} to mean the striation marks made by a screwdriver, {\it scan} to mean the scan obtained for each mark in 3D, {\it signature} to mean the 2D signal we extracted computationally from a mark, {\it replicate} to mean one of the repeated marks made by a single tool at the same condition (e.g., angle and direction), and {\it condition} to mean a combination of angle and direction at which a mark is made.

\section{Previous work}\label{sec:previouswork}

Specifically in non-firearm toolmarks, researchers have scanned toolmarks in 3D and generated algorithms to analyze them since 2001. \citet{geradts} used structured light to capture marks, and then used the variance of gray values to compare signatures. \citet{faden} used surface profilometry, and then separated the signatures into small sections to find the sections with highest cross-correlation. They found that although toolmarks made by the same tool at the same angle could be distinguished from those made by different tools, when the angle varied (30, 60, 85), the marks could not be distinguished from toolmarks made with different tools. \citet{bachrach} used confocal microscopy, and then global relative distance to compare marks made by screwdrivers and tongue-and-groove pliers. They found that marks made by the same tool at different angles differ significantly and equal angles may be required to determine whether two marks are made by the same tool. \citet{chumbley2010} use surface profilometry, and cross-correlation, to find that two toolmarks can only be identified as being created with the same tool if the angle of attack was similar. 

\citet{baiker2014} has a number of interesting contributions. They use global cross correlation to compare marks made at five angles between 15-75 degrees from the normal. Their method has high discriminatory power even at different angles of attack. They find that their automated algorithm beats human examiners in terms of false positives and the humans beat the algorithm in terms of false negatives. They also show that relying on 3D data is better than relying on 2D data, and they find evidence that although their method was trained on certain screwdrivers, it could be used to compare other tools as well. \citet{baiker2015} find that lead preserves fine details in the striations, that it is advantageous to push the tool instead of pulling during toolmark creation for angles of attack above 45 degrees, and that toolmarks should be created as shallow as possible in the substrate material. \cite{garcia2017influence} study axial rotation. Several other projects (\citet{chumbley2009, chumbley2013, grieve, hadlermorris2018, gambino, petraco2012}, to name a few) propose new machine learning and other methods to study 3D marks. 

\citet{hadlermorris2018} developed a method to compare two toolmarks automatically using U statistics. This updates a method from \citet{chumbley2010}, which itself updated a method from \citet{baldwin2004}. The method works by first finding the segments of toolmarks that have the highest correlation (optimization), and aligning the mark based on those segments, and then checking whether the other segments, systematically chosen, also have high correlations (validation). A U statistic (which they call Chumbley U-Statistic) is then calculated based on the correlations of the validation step, and this is the similarity measure. \citet{hadlermorris2018} find separation of the histograms of known match and known non-match pairs in terms of their U statistics. A series of papers \citep{chumbley2009, chumbley2013, chumbley2017} have contributed to this discussion.

Furthermore, \citet{hadlermorris2018} designed an R package\texttt{toolmaRk}\citep{toolmarkpackage}, along with one of our authors, to implement this method. The similarities between our methods are that they align their marks based on selecting segments that have the highest correlation between the two marks. Then they check the remaining segments to see if those also have high correlation. The difference is that they use U statistics for similarity, we use the cross-correlation function.


From the previous work, we learn that classification is improved with 3D marks over 2D marks, and angle of attack can affect the marks. There are different classification approaches that use a variety of scores. We choose to use 3D scans, test for different angles of attack, include direction of mark generation as well, and use a simple similarity score (ccf) to ease with interpretability. There is no clustering analysis to determine what signals should be grouped together in a data-based way a priori, thus we choose to include this. Our hypothesis is that we will learn that toolmarks made at different angles and directions do cluster together, and thus there is hope for the classification of toolmarks despite being made in different ways. Furthermore, there is no likelihood ratio (LR) as far as we are aware. Our method to estimate LRs will help examiners provide their results as a numerical or verbal LR, as recommended \citep{enfsi}.

\section{Data generation}\label{sec:data}

\subsection{Experimental design}

\begin{figure}[htbp!]
   \centering
   \includegraphics[width=.4\textwidth]{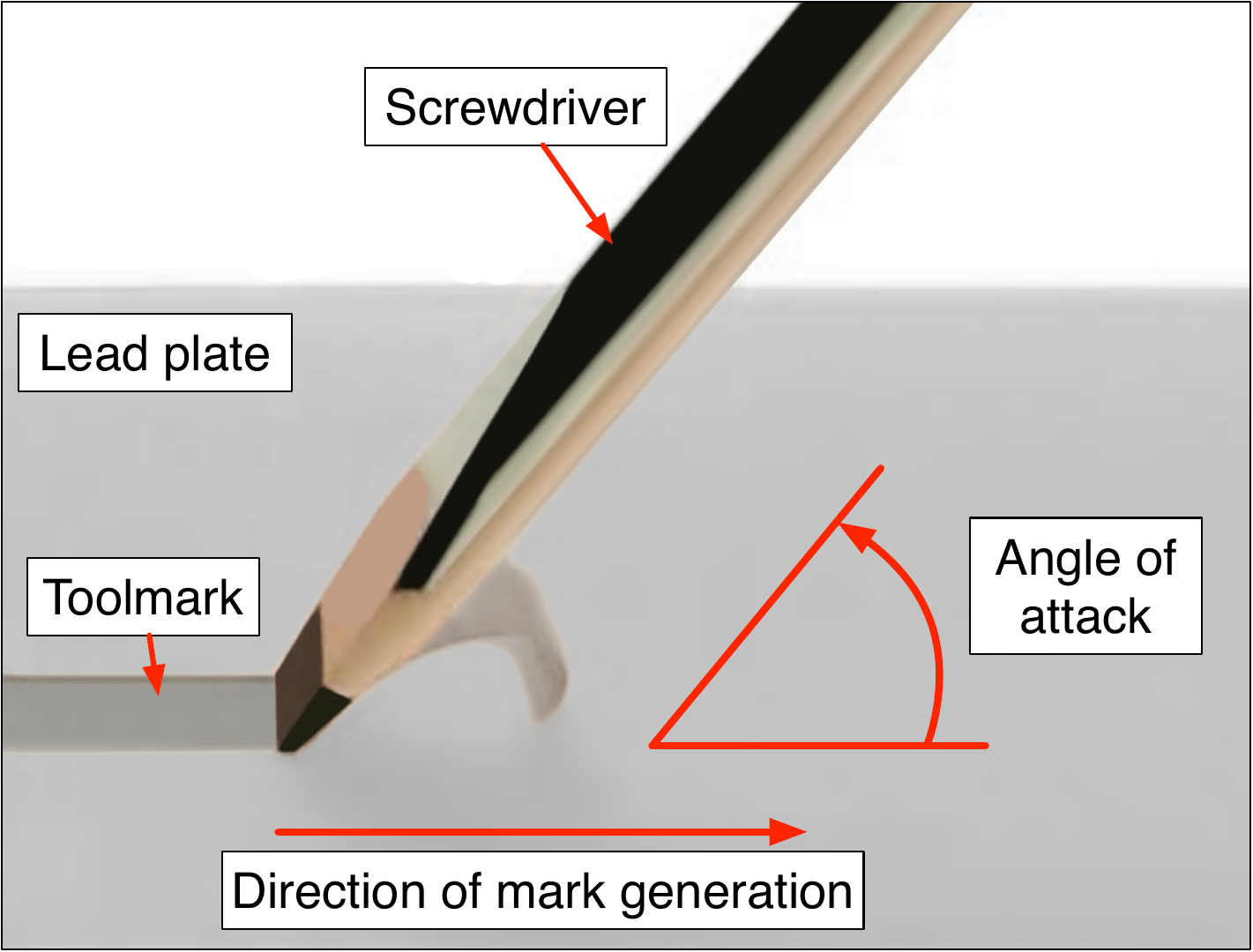}
   \caption{Screwdriver tip generating a striated toolmark on the substrate material. This toolmark is made at a 50 degree angle of attack and in the ``pulling'' direction. Image adapted from \cite{garcia2017influence}.}
   \label{fig:tip}
\end{figure}

We use a factorial design for the toolmarks generation, to allow for the study of the variability within source and between sources of marks, and how this changes as the conditions of of angle of attack and direction of mark generation change. Figure \ref{fig:tip} shows how we define the angle (rising from the surface, so that when the screwdriver is perpendicular to the plate it is at 90 degrees) and direction of mark generation (pulling here means to the right and pushing to the left).

\begin{table}[h] 
\centering
\caption{Experimental design. For each of the three experiments (1, 2, and 3), we altered a different variable (tool, angle, direction) and generated replicates at each condition.}
\resizebox{1\textwidth}{!}{
\begin{tabular}{ccccccc} 
\toprule
Experiment              & Num.\,of tools (Size) & Sides    & Angles      & Directions         & Replicates & Num. of marks  \\
\midrule
1: Tool           & 20 (S)          & A, B & 80         & Pull                 & 8         & 320              \\
2: Angle          & 3 (L)           & A, B & 60, 70, 80 & Pull                 & 8         & 144             \\
3: Direction      & 3 (S)           & A, B & 80         & Pull, push           & 8         & 96               \\
\bottomrule
\end{tabular}}
\label{tab:experimentaldesign}
\end{table}

Table \ref{tab:experimentaldesign} shows the combination of conditions for each. The factors are the angle and direction. We generate three sets of marks, which we call experiments, for a total of 560 marks. As can be seen in Table \ref{tab:experimentaldesign}, we generate eight replicates per tool-side under the same conditions. Note that we used large screwdrivers for experiment 2, since varying angle required long screwdrivers to reach the lead plate. Experiments 1, 2, and 3 allow us to study the variability between marks made by the same tool, since there are eight replicates made under each condition. The angles are 60, 70, and 80 degrees with respect to the lead surface. To clarify, an angle of attack of 90 degrees means that the shaft of the screwdriver is perpendicular to the surface of the substrate. The directions are pushing and pulling, where pulling refers to a negative rake angle. For example, Figure \ref{fig:tool1aligned} shows the marks made by a screwdriver, with eight replicates. Experiment 1 includes marks produced at a constant angle and direction. Experiments 2 and 3 focus on the variability between marks made at different angles and directions. Figures \ref{fig:tool1aligned-angle} and \ref{fig:tool1aligned-direction} show that the marks vary some when made at different angles and directions. We quantify these differences in the Methods and Results sections (\ref{sec:methods} and \ref{sec:results}, respectively).

\subsection{Materials}\label{sec:materials}

\begin{figure}[h!]
\hfill
\subfigure[Slotted screwdriver from Klein tools, consecutively manufactured by size.]
{\includegraphics[width=.4\textwidth]{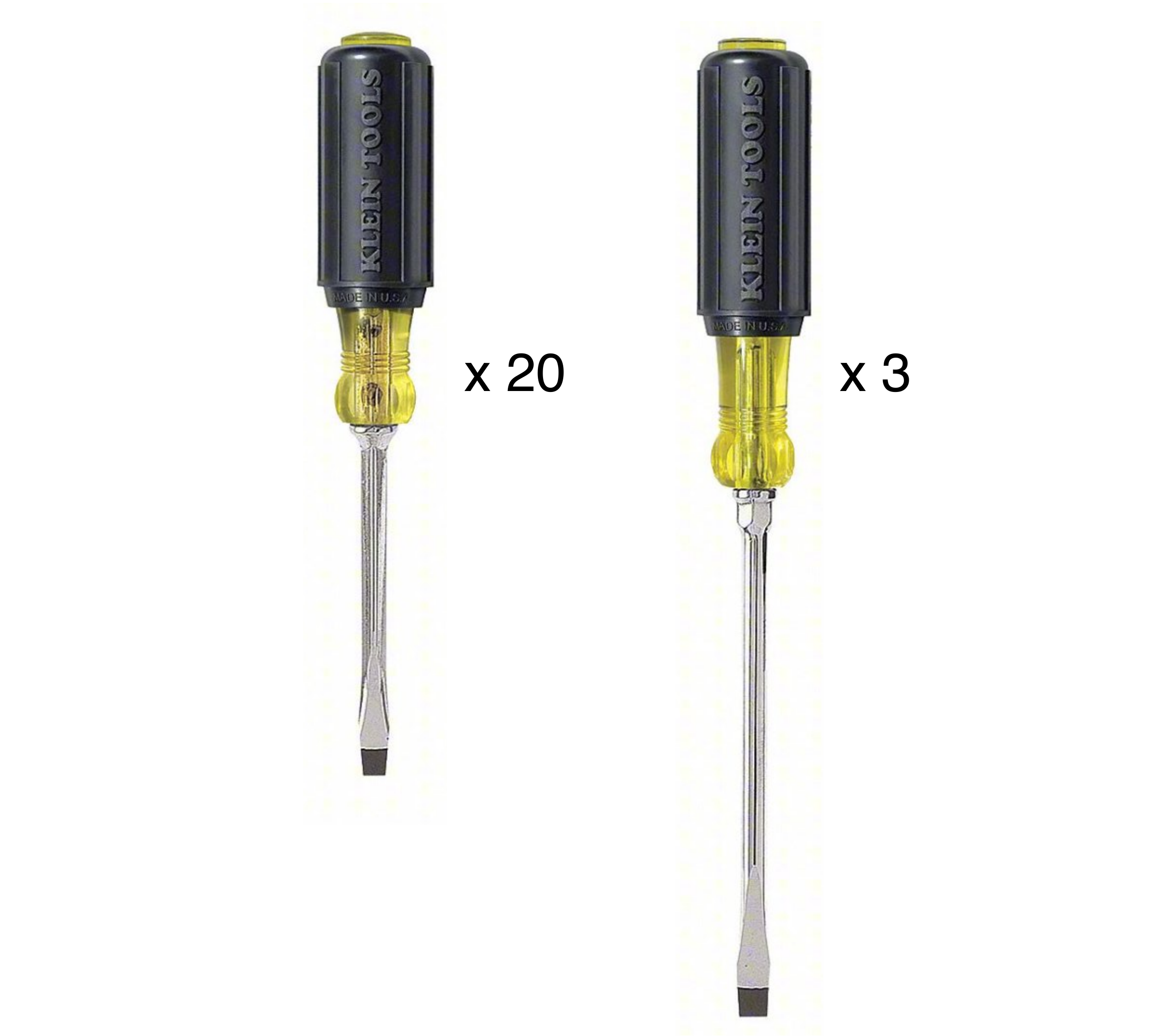}}
\quad
\subfigure[Mechanical rig used to generate striation screwdriver toolmarks in a controlled way.]
{\includegraphics[width=.55\textwidth]{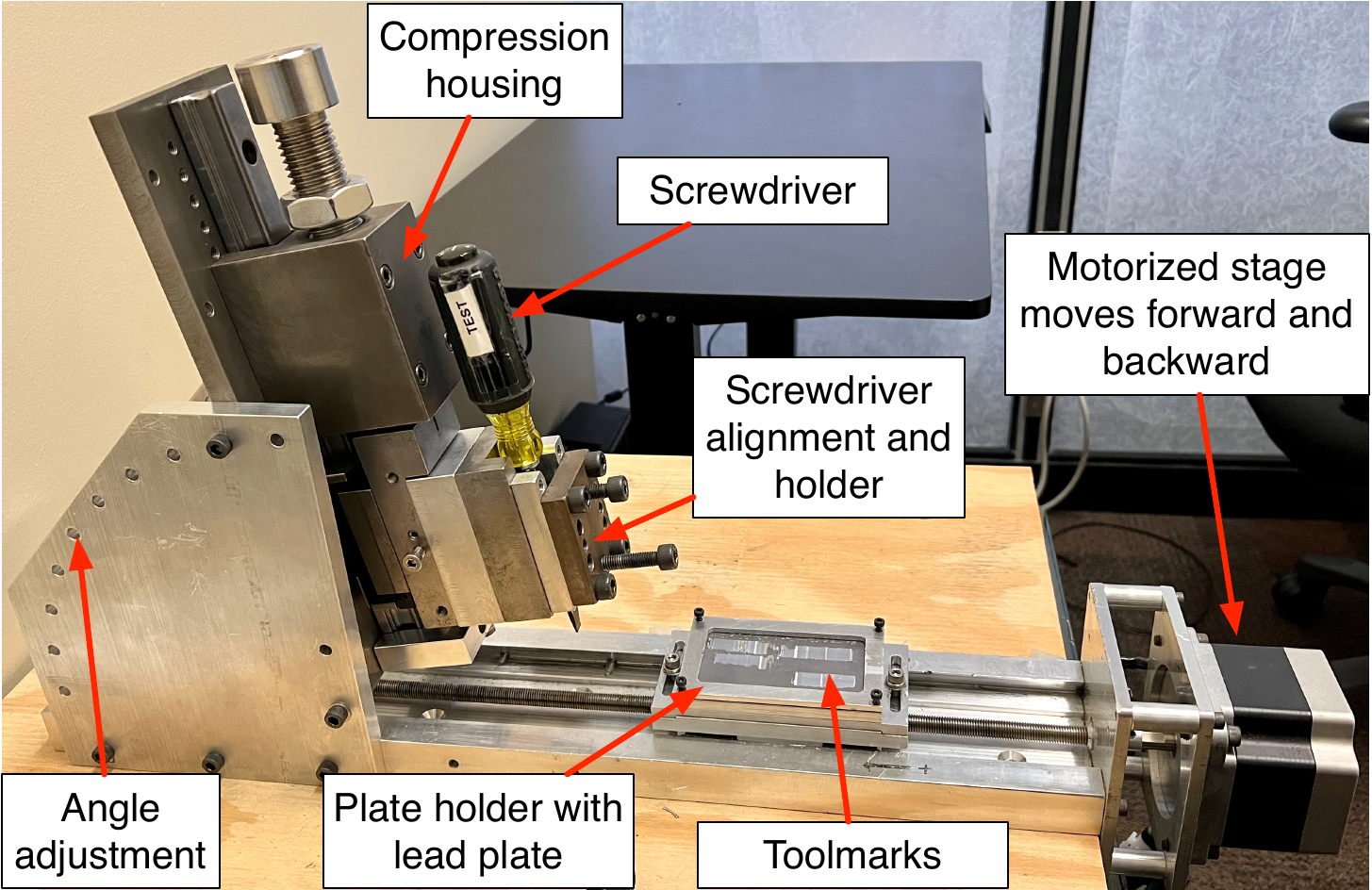}}
\hfill
\subfigure[Striation toolmarks made with a screwdriver on a lead plate, using the mechanical rig.]
{\includegraphics[width=.45\textwidth]{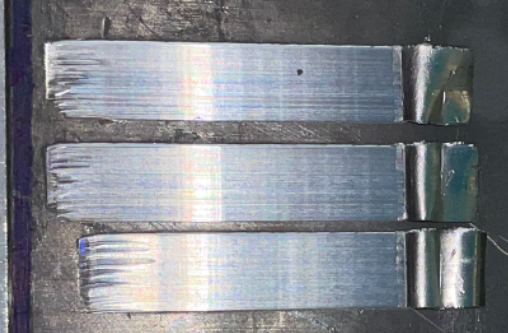}}
\quad
\subfigure[GelSight handheld Mobile 1.0X scanner and tablet with GelSight software used to obtain the 3D scans of the toolmarks as shown in Figure~\ref{fig:process}a.]
{\includegraphics[width=.45\textwidth]{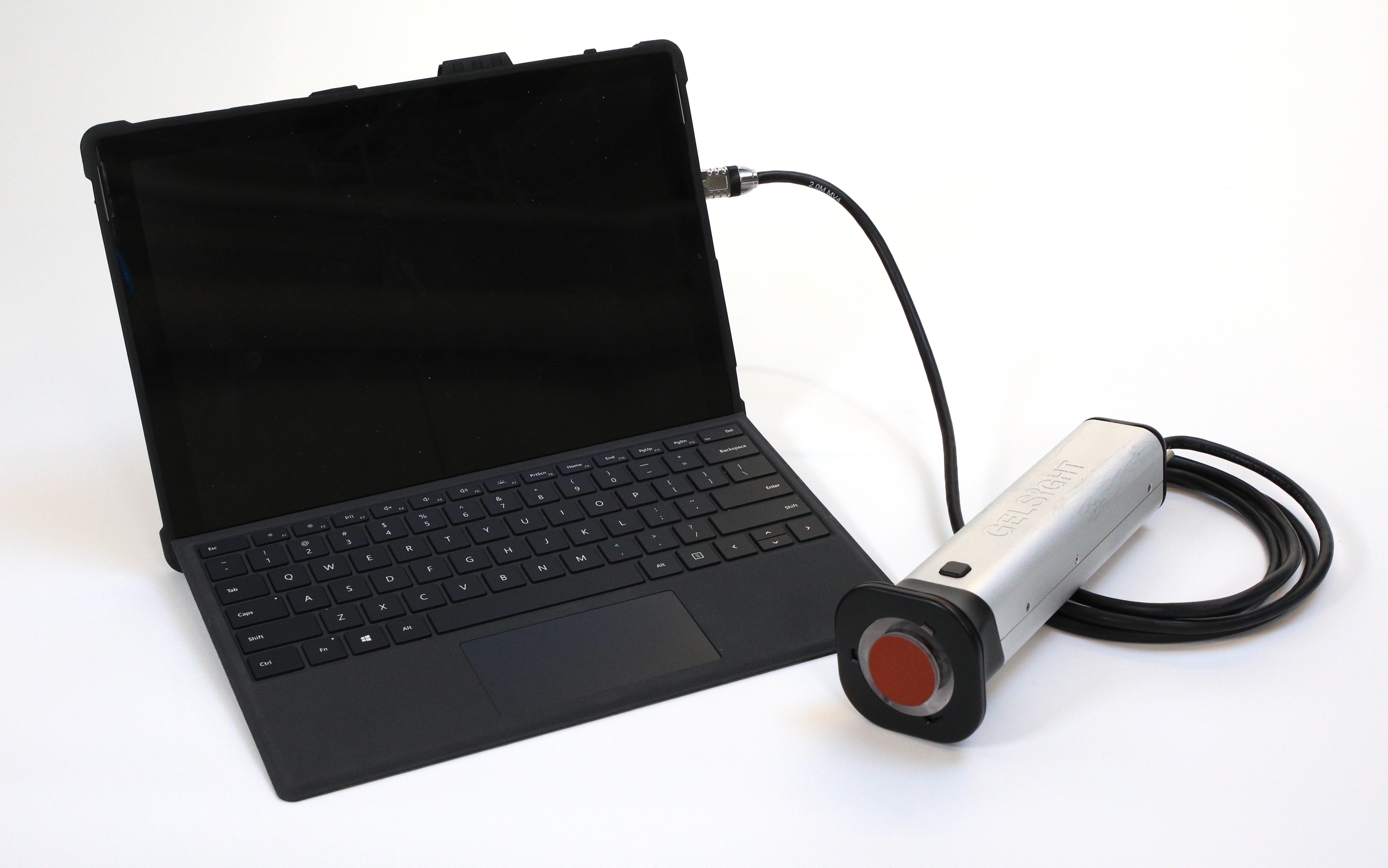}}
\hfill
\caption{Materials for generating toolmarks and scanning them in 3D.}
\label{fig:materials}
\end{figure}

The materials for the experiments are 20 small slotted consecutively manufactured screwdrivers (Figure \ref{fig:materials}a) and 3 large slotted screwdrivers. Small screwdrivers 1-20 were consecutively manufactured in one set, and large screwdrivers 1-3 consecutively manufactured in another. This allows us to test a challenging scenario in which different sources are potentially very similar to each other due to manufacturing processes. It is likely that sub-class characteristics, such as a striation created by the manufacturing process, could make the toolmarks from different sources similar to each other \citep{nichols1, nichols2}. 

Regarding the screwdriver manufacturing process at the Klein tools factory, an automated Siepmann grinding wheel is used to form the final tip geometry in the manufacturing process. Since the grinding wheel is used independently for each side of the screwdriver, the marks on the two sides are likely very different from each other. Klein tools, the manufacturer, sells tools to a variety of suppliers, including Home Depot, where they may be purchased for amounts between \$10 and \$15. We select flat-head screwdrivers for our study because they can be used to generate striation marks on a flat surface, unlike other tools such as wire cutters that have multiple surfaces that interact with each other. 

To generate toolmarks in a controlled and replicable way, we use a mechanical rig (Figure \ref{fig:materials}b), which we obtained from the authors of \cite{zhengetalrig} and modified in the Manufacturing \& Fabrication Services shop the University of Pennsylvania Department of Mechanical Engineering \& Applied Mechanics. The rig has a Velmex motorized slide that controls the direction and speed of the plate to allow for the generation of replicable marks. 

We used flat lead plates as the substrate for the screwdriver striation marks (Figure \ref{fig:materials}c). We generate the toolmarks on lead plates because we find that lead preserves fine details in the striations~\citep{baiker2015}, and it allows us to create smooth marks without high forces, which can make the slide motor stop and create jitter effects. We made test marks in machinable wax, copper, and aluminum. None of these materials captured the entire screwdriver mark as well as lead, so we chose lead. As this is a foundational study, it is important to observe the mark made by the entire screwdriver tip. However, future work could study how changing the material affects the marks.

To scan the toolmarks in 3D, we use a handheld scanner sold by GelSight, called the Mobile 1.0X handheld instrument (Figure \ref{fig:materials}d). The GelSight instrument's resolution~\citep{gelsight} in the $x$-$y$ (horizontal) plane is 3.45 microns, and the accuracy in the $z$ (vertical) direction is 4 microns. This scanner is hand-held and powered by a tablet, which could make it easy to use at crime scenes. It cannot be used to scan very deep marks (it has been tested up to 90 microns in depth~\citep{gelsight} or sharp-angled surfaces that could break the gel's surface. However, it is well-suited for scanning striation marks like the ones in this experiment and likely for many marks found at crime scenes. The scans are given by GelSight in STL format, and we converted them to x3p format, which is more common in forensic statistic analysis. The data is publicly available in [forthcoming].

\begin{figure}[htbp!]
\subfigure[Step 1: A rendering of a 3D toolmark scan, obtained with the GelSight handheld Mobile 1.0X scanner.]
{\includegraphics[width=.45\textwidth]{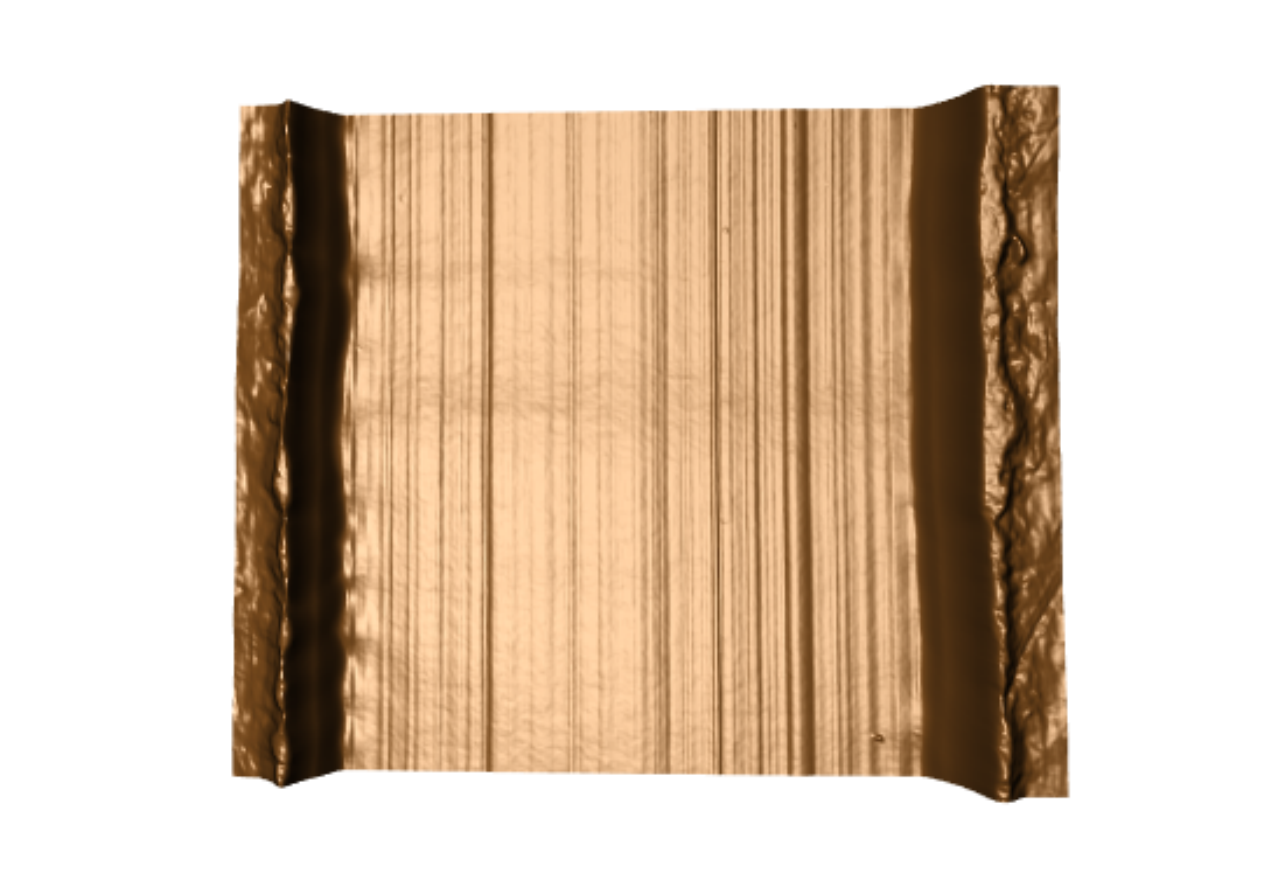}}\quad 
\subfigure[Step 2: The black line is the location of the cross section.]
{\includegraphics[width=.45\textwidth]{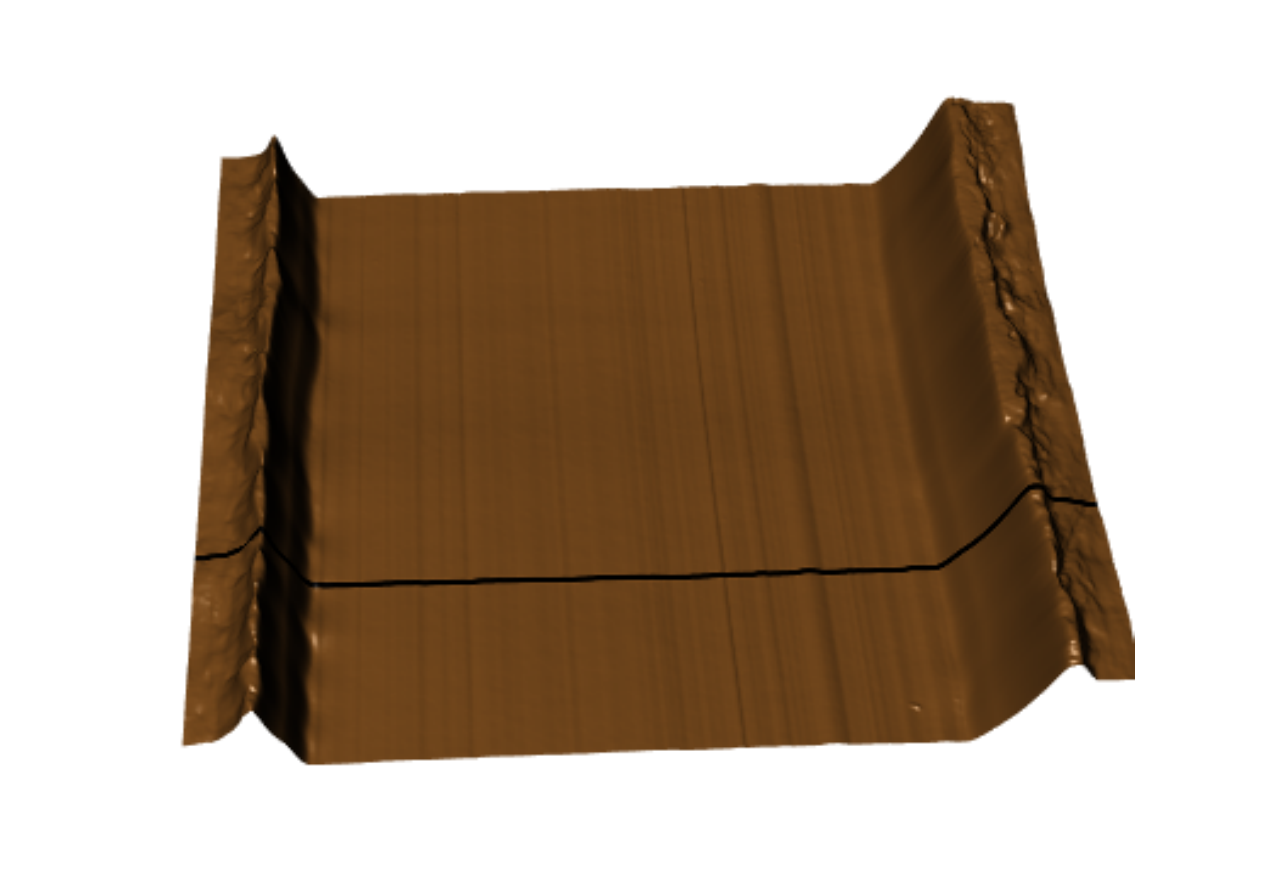}}
\subfigure[Step 3: The height profile corresponding to the cross section. The blue vertical lines are selected manually as the edges of the mark for cropping.]
{\includegraphics[width=.45\textwidth]{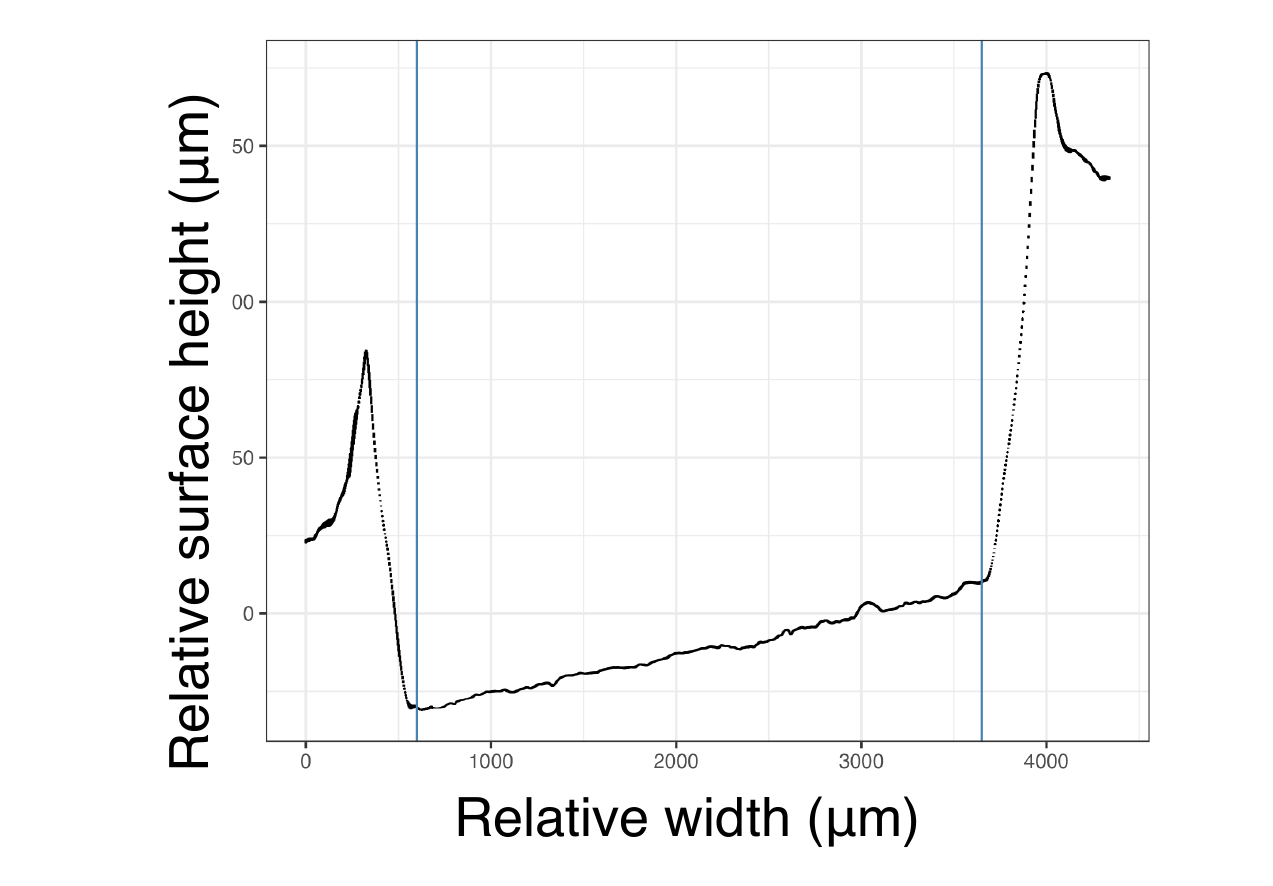}}\quad
\subfigure[Step 4: The cropped signal in black, and a blue curve showing Gaussian smoothing.]
{\includegraphics[width=.45\textwidth]{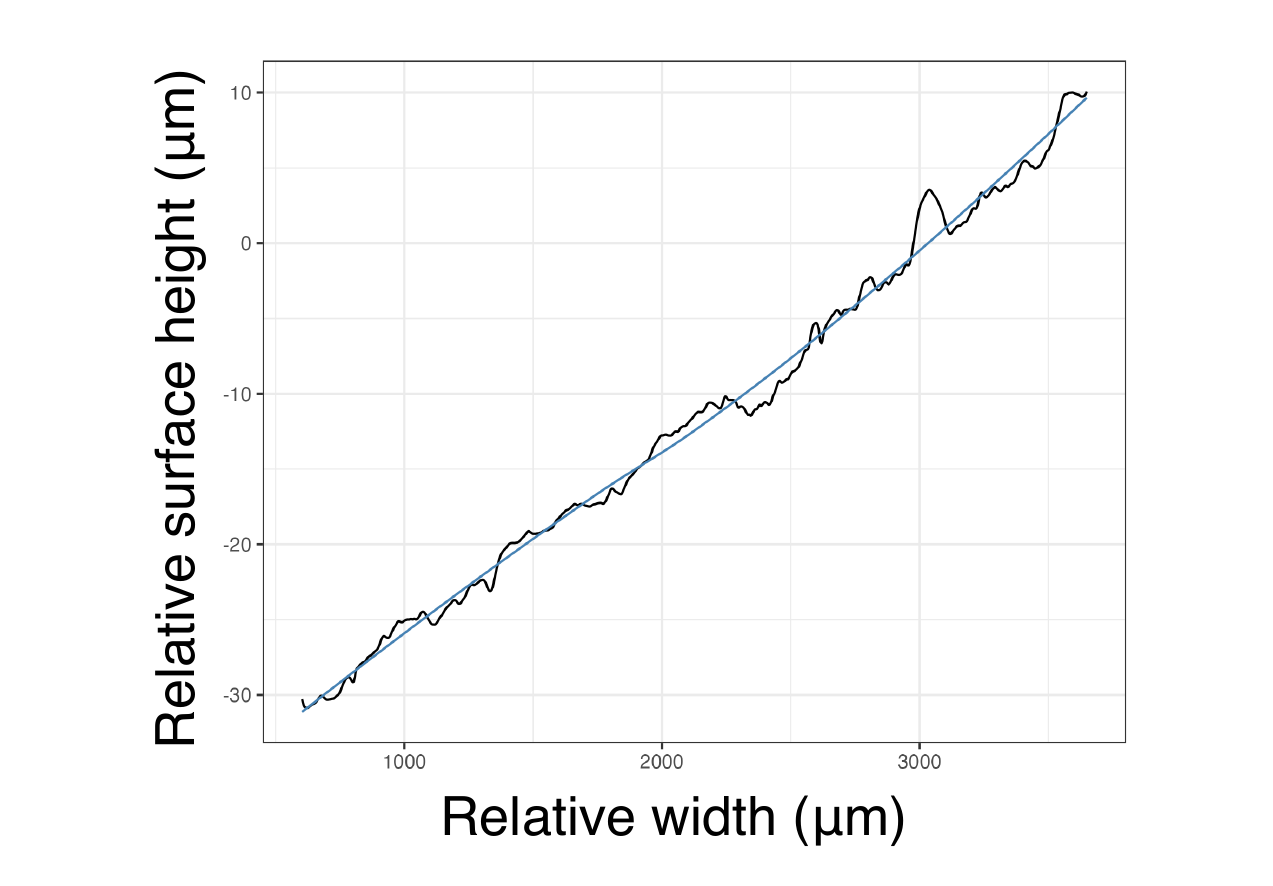}}\hfill \centering
\subfigure[Step 5: Extracted signal. This is the residual between the black signal and the blue curve in (d). For illustration, another signal from a replicate mark is overlaid in the dotted profile.]
{\includegraphics[width=.7\textwidth]{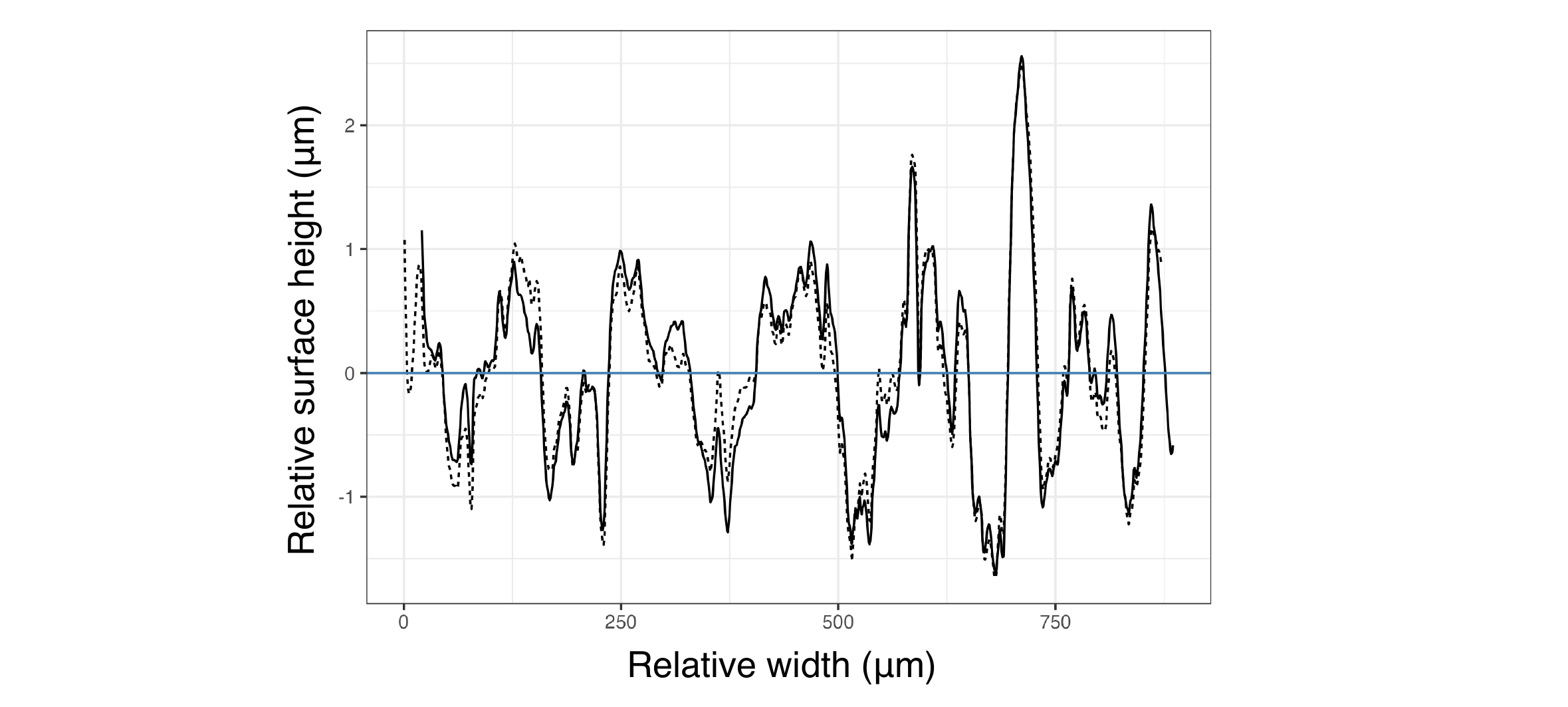}}
\caption{Steps to extract the signals from the 3D toolmark scans.}
\label{fig:process}
\end{figure}

\subsection{Signal extraction}\label{sec:sigextraction}

The process to extract the digital signals from the 3D scans is shown in Figures~\ref{fig:process}a-\ref{fig:process}e. After obtaining the 3D scan from the Gelsight instrument (Figure~\ref{fig:process}a), we select a cross section in the middle of the scan (Figure~\ref{fig:process}b). The reason for extracting the 2D signal from the 3D scan is that in our experimental setup, the striation marks are consistent throughout the toolmark. We verified this by taking several cross sections throughout the toolmark and comparing them. The information about the mark made by the screwdriver tip is contained in the relative depths of the striation marks. Going from 3D to 2D in this way is a compression that preserves the relevant information of the mark made by the screwdriver tip. The 2D signal we extract is in one way comparable to a 2D light microscope scan of the toolmark, as is used by many forensic examiners, since it contains information about the striation marks. However, our signal has the advantage of containing precise relative depth information since it is a cross-section of the toolmark. 

To obtain a cross-section of the toolmark, we use the R package \texttt{bulletxtrctr} \citep{bulletxtrctr}. We select a point in the vertical middle of the mark. We crop the edges of the profile manually (Figure~\ref{fig:process}c), and apply Gaussian smoothing to the data to model macro structures (Figure~\ref{fig:process}d), such as the plate's shape and any scanning-specific trends. For Gaussian smoothing, we used the R loess function with span parameter set at 75\%. We define the signal of the screwdriver side as the difference between the profile's height values and the fitted structure. In other words, the signal is the residual between the profile height and the fitted smoothed signal, which removes the macro structure and normalizes the signal so it is on a flat horizontal surface. Figure~\ref{fig:process}e shows the signals extracted from two different replicate marks, made by the same source.

\subsection{Data}\label{sec:finaldata}

\begin{figure}[htbp!]
   \centering
   \includegraphics[width=.8\textwidth]{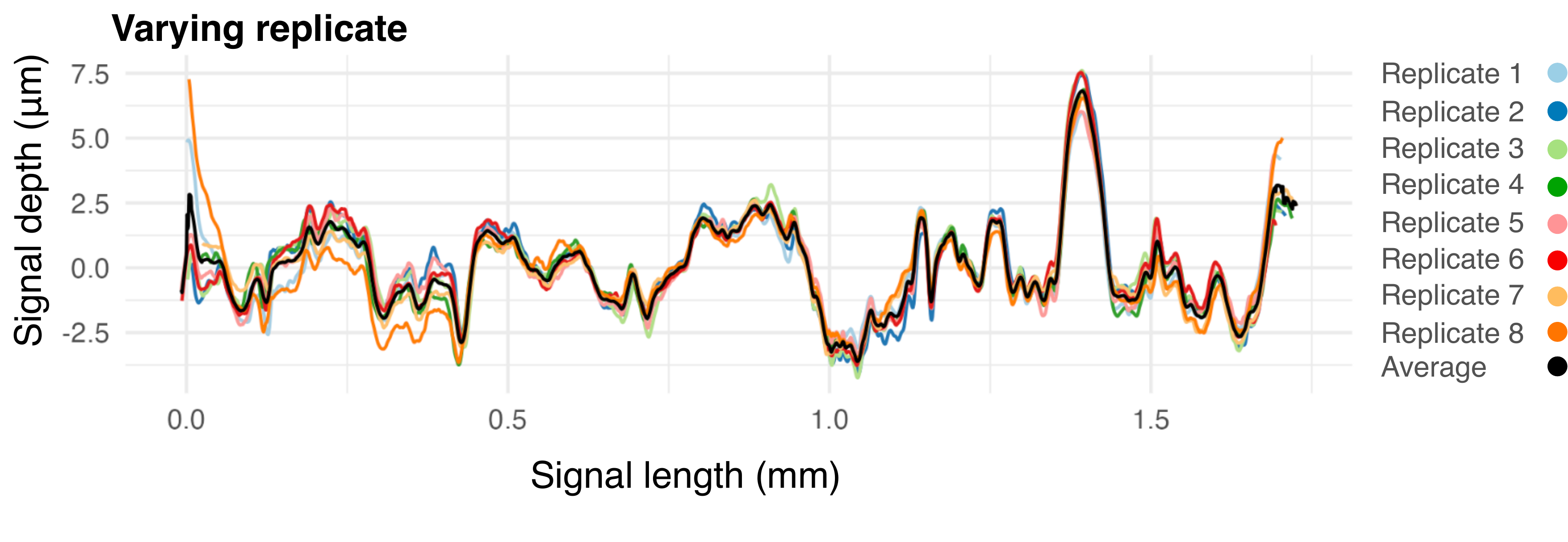}
   \caption{Replicate signals from a single source (small tool 1), at a fixed angle of attack (80) and direction of tool generation (pull). The black signal is the average of the rest.}
   \label{fig:tool1aligned}
\end{figure}

\begin{figure}[htbp!]
   \centering
   \includegraphics[width=.8\textwidth]{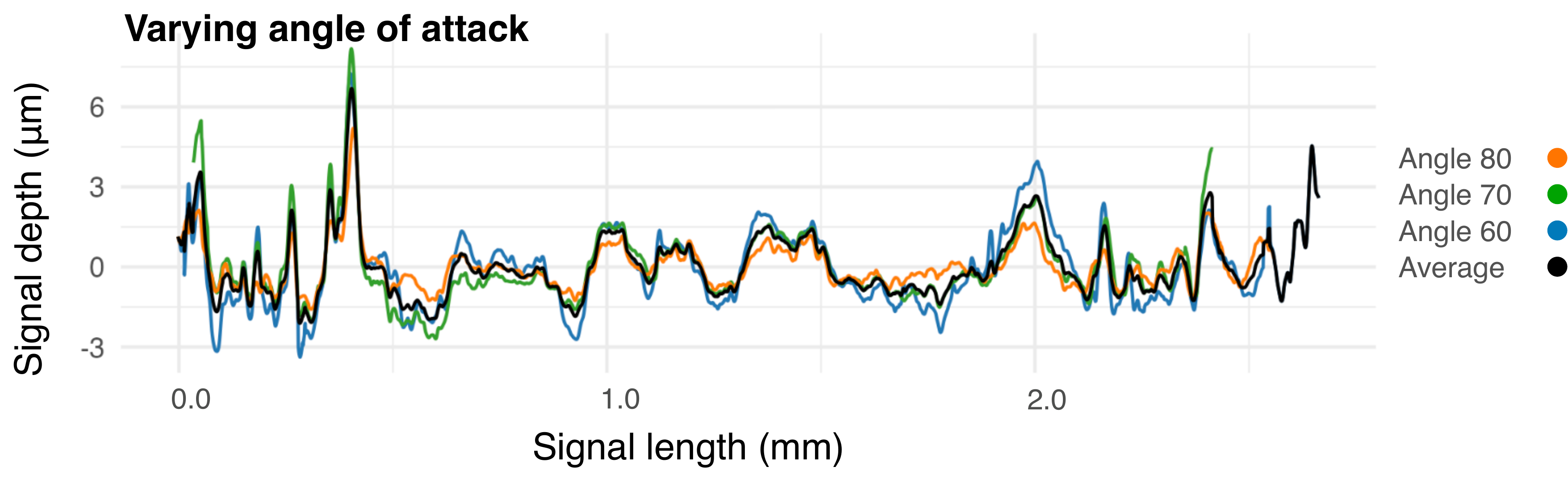} 
   \caption{Averaged replicate signals from a single source (large tool 1), at three angles of attack (60, 70, 80), at a fixed direction (pull). Eight replicates made at each angle were averaged across angle, so only three signals are shown. The black signal is the average of the other three curves. Note that these signals are wider than those of experiments 1 and 2 because they were made with larger screwdrivers.}
   \label{fig:tool1aligned-angle}
\end{figure}

\begin{figure}[htbp!]
   \centering
   \includegraphics[width=.8\textwidth]{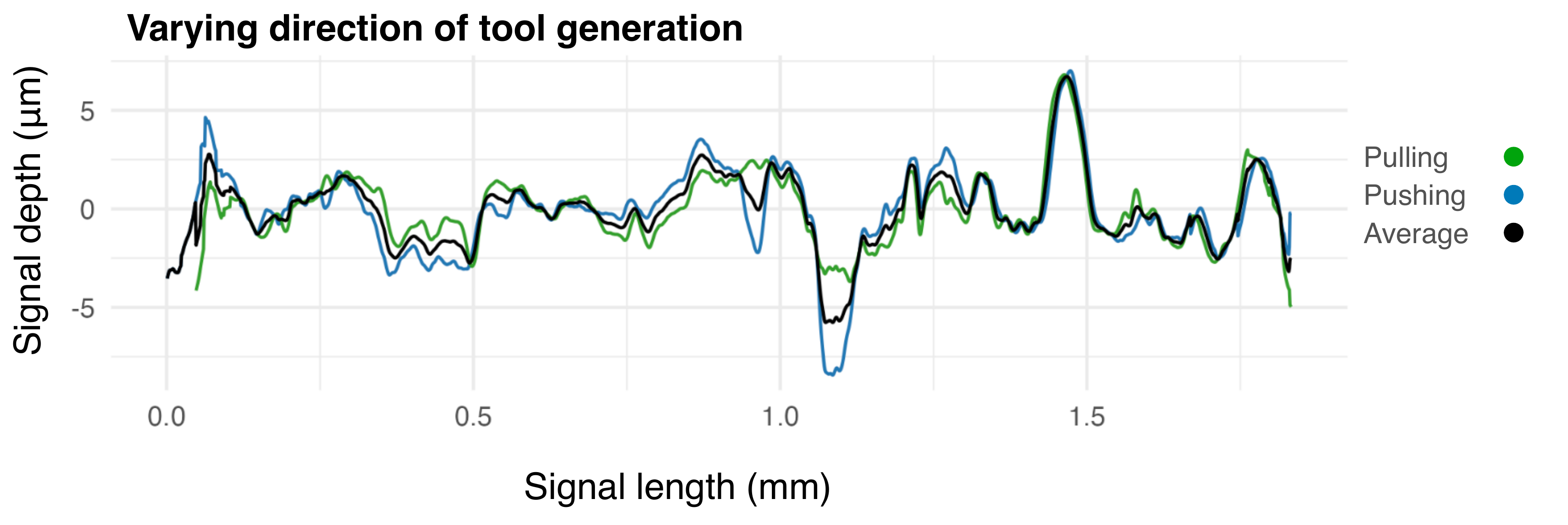} 
   \caption{Averaged replicate signals from a single source (small tool 1), at two directions of tool generation (push and pull), at a fixed angle (80). Eight replicates made at each direction were averaged across direction, so only two signals are shown. The black signal is the average of the other two curves.}
      \label{fig:tool1aligned-direction}
\end{figure}

Finally, we align the signals. Figures~\ref{fig:tool1aligned}-\ref{fig:tool1aligned-direction} show examples of signals from our dataset.
Figure~\ref{fig:tool1aligned} shows replicate signals from small tool 1, side A, at a fixed angle and direction. Qualitatively, the replicates look quite similar to each other. Figure~\ref{fig:tool1aligned-angle} shows averaged replicate signals (one for each average of eight replicates), made by large tool 1 at different angles, and fixed direction. Figure~\ref{fig:tool1aligned-direction} shows averaged replicate signals made by small tool 1 at different directions, and fixed angle. Note that \ref{fig:tool1aligned} and \ref{fig:tool1aligned-direction} are similar because they are made with the same tool. We did not make the varying-angle marks with small tool 1 because the rig required longer tools to reach the lead plate at varying angles. It also required additional force at lower angles of attack. We leave studying these effects for future research.

\section{Methods}\label{sec:methods}

\subsection{Method 1: Similarity matrices and clustering to studying variability by source, angle, and direction}

We first cluster the signals to study the variability within tool and the variability between tools, both at a fixed setting and when varying angle and direction. We use clustering as an exploratory step that allows us to know which pairs we should consider same-source or different-source in a data-driven way. In other words, we do not assume a priori that the marks made by the same tool are more similar to each other than marks made by different tools. We allow the clustering algorithm to select which sets of marks should be considered part of the same group. This step could be particularly useful in testing whether new factors, such as angle of rotation, substrate material, and force, and interactions of factors, generate marks that are so different from each other that they could be considered to have been made by a different tool.

For alignment, we use a sliding window approach (sometimes called ``registration'') in which signals are compared in pairs by sliding one over the other, to find the lag that produces the maximum correlation between the signals. See \cite{hare2017algorithmic} for a more detailed description of this process. We then use similarity measures that are extracted from pairwise aligned signals. 

Clustering allows us to see whether varying angle and direction leads to great differences in the signals, an effect that could lead toolmark comparisons to be very challenging since for each tool examiners might have to consider a great (possibly infinite) number of marks. We cannot simply assume that marks made at different angles and directions, by the same tool-side, can be considered same- or different-source, since it is possible that at different settings, marks look extremely different from each other. Signals made at angles of attack greater than 15 degrees difference could not be considered same-source because they were too different from each other \citep{hadlermorris2018}. This clustering test is a data-driven method to determine whether we should consider pairs of marks to be same-source or different-source. It also helps compare the within-tool variability across replicates to the between-tool variability, for each tool. Of course, this is limited to the training data that we produced. Thus, clustering is a preliminary test for us to select which pairs to include in the densities in the rest of our methodology.

For clustering, we use the Partitioning Around Medoids (PAM) clustering method, also called the k-medoids clustering method, which, like k-means, partitions the dataset into groups and minimizes the distance between points and their cluster centers. Unlike k-means, k-medoids selects actual data points as centers, making the clusters easier to interpret. K-medoids can also work with any dissimilarity measure, unlike k-means, which typically requires Euclidean distance. This approach is more robust to noise and outliers because it minimizes pairwise dissimilarities instead of squared Euclidean distances.

Formally, for each experiment, given the similarity matrix $S$, the next step is to cluster the toolmarks into different groups based on $S$. In general the goal of clustering algorithms is to partition data into different groups such that data within a group are more similar than data from different groups. For Euclidean data, $k$-means is a popular clustering method in which the mean of the data within one group meaningfully represents the ``center'' of the cluster. However, this property does not hold for non-Euclidean data where the similarity measure is arbitrary \citep{schubert2019faster}. Due to the nature of the toolmarks similarity, we use the partition around medoids (PAM) algorithm \citep{dodge1987introduction, kaufman2009finding} which is a generalization of $k$-means to non-Euclidean data. The goal of the algorithm is to minimize the average dissimilarity of objects to their closest selected object. The medoid of a set $C$ is defined as the object with the smallest sum of dissimilarities (or, equivalently, smallest average) to all other objects in the set hence it can be viewed as a representative of the objects in this cluster. In summary, the PAM algorithm searches for $k$ representative objects in the given dataset (denoted as $k$ medoids) and assigns each point to the closest medoid to create clusters. The objective is to minimize the sum of dissimilarities between the objects within a cluster and the center of the same cluster (i.e, medoid).

Before calculating the clustering result for each experiment, we need to select the number of clusters $k$ for each experimental setting, that is, to identify the optimal number of clusters to include. To this end, we need a measure for how ``good" a clustering algorithm is. The Silhouette score \citep{rousseeuw1987silhouettes} evaluates clustering performance based on the pairwise difference between within-cluster and between-cluster distances. Given a similarity matrix $S$ and a cluster $C_i$, for each datum $n \in C_i$, we first calculate,

\begin{align}
    \begin{split}
        a(n) &= \frac{1}{\vert C_i\vert-1}\sum_{m\in C_i, m\neq n} S_{mn}\\
        b(n) &= \min_{j\neq i} \frac{1}{\vert C_j\vert}\sum_{m\in C_j} S_{mn},
    \end{split}
\end{align}
as the mean within-cluster distance and smallest between-cluster distance. Here, we denote $\vert C_i\vert$ as the size of the $i$-th cluster for $i=1, 2, \dots, k$. The Silhouette score for datum $n \in C_i$ is then defined as \begin{equation}
    \label{eq:silhouette_score}
    s(n)=\frac{b(n)-a(n)}{\max\{a(n), b(n)\}}
\end{equation}
if $\vert C_i\vert>1$ and $s(n)=0$ otherwise. By definition, $-1\leq s(n)\leq 1$, and we define the Silhouette score for the clustering method as the average Silhouette score across all samples.

We then vary the number of clusters and apply PAM clustering for each possible $k$. For the clustering result at each given cluster number, we calculate the average Silhouette scores across all samples. We then choose the cluster number that maximizes the Silhouette score. Intuitively, this corresponds to the cluster number that yields the ``best partition'' of the data.

\subsection{Method 2: Known-match and known-non-match densities to classify same- and different-source}

Second, we plot a density of similarity scores observed among known matches and a density of scores observed among known non-matches. We seek to find 1) whether the densities are separated such that there is a small overlap between them, and if so 2) the threshold, i.e., where the two densities cross in terms of similarity. We use the threshold to classify between whether there is more support for the evidence given the prosecution or the defense hypothesis. This threshold helps later to test the performance of the classifier. We then fit distributions on the similarity densities to allow for the estimation of likelihood ratios for new pairs of toolmarks.

\subsection{Method 3: Score-based likelihood ratio to provide probabilistic interpretation}

Third, we introduce a score-based likelihood-ratio approach to make forensic toolmark comparisons. In a criminal case, forensic examiners analyze the evidence and present their findings to the trier of fact, who combines all the information presented in the case to deliver a final decision about the defendant's guilt. Under a probabilistic framework, the trier of fact compares two propositions referred to as the prosecution hypothesis ($H_p$) and the defense hypothesis ($H_d$) conditional on the evidence observed \citep{aitken}. Applying the ratio form of Bayes' theorem, the trier of fact's task is to estimate,

\begin{equation}
\underbrace{\frac{P(H_p | E)}{P(H_d | E)}}_{\text{Posterior odds}} = \underbrace{{\frac{P(E | H_p)}{P( E | H_d )}}}_{ \text{Likelihood ratio} }\underbrace{ \frac{P(H_p )}{P(H_d )}}_{\text{Prior odds}}.
\label{eq:br}
\end{equation}
In other words, the trier or fact's prior beliefs regarding the hypotheses are updated via a likelihood ratio (sometimes called a Bayes' factor). Forensic experts are advised to present their findings as a likelihood ratio by scientific and professional organizations \citep{enfsi}. 

In the case of forensic toolmarks, experts may be presented with a pair of questioned marks as evidence $E = (E_x, E_y)$ and asked to evaluate if a common tool produced the two marks. Under the common-source framework \citep{ommen}, we can state the propositions as $H_p:$ Marks $E_x$ and $E_y$ were made by the same unknown tool, and $H_d:$ Marks $E_x$ and $E_y$ were made by different unknown tools.

To assess these competing propositions, forensic experts can rely on observed features of the questioned mark. Let $u_i$ denote the features of $E_i (i=x,y)$. If the joint distribution of the features under each of the competing propositions, denoted by $f(u_x,u_y | H_j) (j=d,p)$, is known, the likelihood ratio could be computed, 
\begin{equation}
LR = \frac{f(u_x,u_y | H_p)}{f(u_x,u_y | H_d)}.
\end{equation}
A $LR>1$ indicates that the priors are being updated towards the prosecutor, meaning the evidence supports the prosecutor's proposition, while a $LR<1$ indicates that the priors are being updated towards the defense.

To estimate the joint probability model, researchers use a sample of the background population or reference set composed of information previously collected. Let $A$ denote the reference set, $E^A_{ij}$ an individual item $j(j=1, \dots, n_i)$ from source $i$, $(i=1, \dots, m)$ and $A_{ij}$ the corresponding measurement from item $j$ from source $i$. Note that here, our reference set is the data from the three experiments.

\citep{ommen} express the proposition and the process that generated the data available to the expert as a sampling model. They consider that the reference set $A$ was generated first by randomly sampling $m$ sources from a reference population and, within each source, sampling $n_i$ items. To do this, experts have relied on machine learning comparison metrics and density estimation procedures to construct score-based likelihood ratios. This is the approach we take here.

We use a score-based likelihood ratio approach, as others have done before \citep{carriquiryetal2015, hadlermorris2018, baiker2014, veneri2023ensemble, tai2018fully, baiker2014, hare}. For all the pairs of signals, we plot the known match (KM) and known non-match (KNM) densities of the similarities. To handle the dependencies produced by the replicates, we provide three approaches: averaging correlations across source, a naive method assuming independence, and sampling \citep{veneri2023ensemble}. As a measure of similarity, we use the correlation of the aligned signals, sometimes called the cross-correlation function, transformed to be between 0 and 1. For the remainder of this article, we refer to this transformed correlation as the ccf or similarity score.

Then, we fit probability distributions to the densities to be able to quantify the height of the curve at a certain measure of similarity. The quotient of the height of the KM curve over the height of the KNM curve at a given similarity score yields the value of the likelihood ratio.

The likelihood ratio should be interpreted as follows: If LR is less than 1, then there is support for the defense hypothesis, if it is greater than one then there is support for the prosecution hypothesis, and if it equals one, then there is equal support for both. A likelihood ratio of 20 can be interpreted as the conclusion that it is 20 times more likely to observe this similarity if the toolmarks were made by the same tool than from different tools. One can use likelihood ratios to classify pairs into whether there is more weight on the prosecution's hypothesis or the defense hypothesis, but that is \textbf{not} the same as determining whether the pair is truly made by the same source or different sources, because that would be posterior odds. In order to be used in court, the results of a likelihood ratio analysis need to be reported verbally to the trier of fact. \cite{enfsi} provide guidelines for how to translate numbers to a verbal scale. For example, LR=20 is ``moderate support'' for the prosecution's hypothesis over the defense's hypothesis. This verbal scale is an effective procedure for translating from the result of this quantitative method to a correct qualitative measure that is understandable and correctly interpreted by a lay population.

\section{Results}\label{sec:results}

\subsection{Method 1: Similarity scores and clustering}

For clustering, the first step is to calculate the similarity matrix for each experiment, meaning all the pairwise cross-correlation functions in the data. See the similarity matrices for each experiment Figures~\ref{fig:simmatrix_exp1}, \ref{fig:simmatrix_exp2}, and \ref{fig:simmatrix_exp3}. Qualitatively, note that there is blocking by source, angle, and direction in the similarity matrices. In all three matrices, the orange blocks off the diagonal suggest that there might be false positives if similarity is used to classify between same- and different-source. The gray blocks in the main diagonal suggest that there might be false negatives. A statistical test for whether there really are ``blocks'' or clusters, is by running a clustering algorithm.


\begin{knitrout}
\definecolor{shadecolor}{rgb}{0.969, 0.969, 0.969}\color{fgcolor}\begin{figure}

{\centering \includegraphics[width=\textwidth]{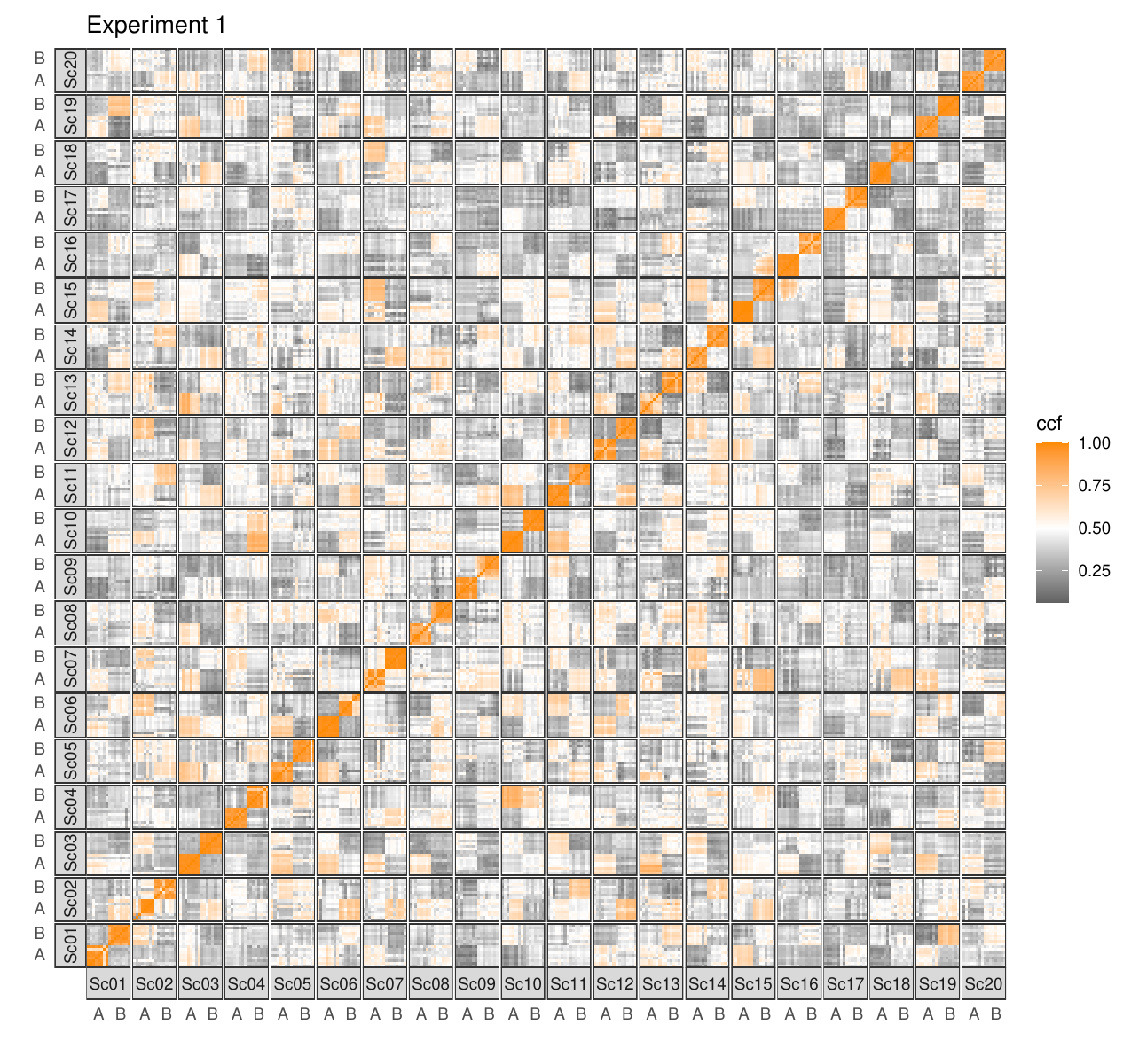} 

}

\caption[Similarity matrix displayed as a heat map for the similarities between toolmarks made by 20 small screwdrivers, sides A and B, with 8 replicates each]{Similarity matrix displayed as a heat map for the similarities between toolmarks made by 20 small screwdrivers, sides A and B, with 8 replicates each. Each tiny square corresponds to the pairwise similarity (ccf) between two toolmarks.}\label{fig:simmatrix_exp1}
\end{figure}

\end{knitrout}


\begin{knitrout}
\definecolor{shadecolor}{rgb}{0.969, 0.969, 0.969}\color{fgcolor}\begin{figure}

{\centering \includegraphics[width=\textwidth]{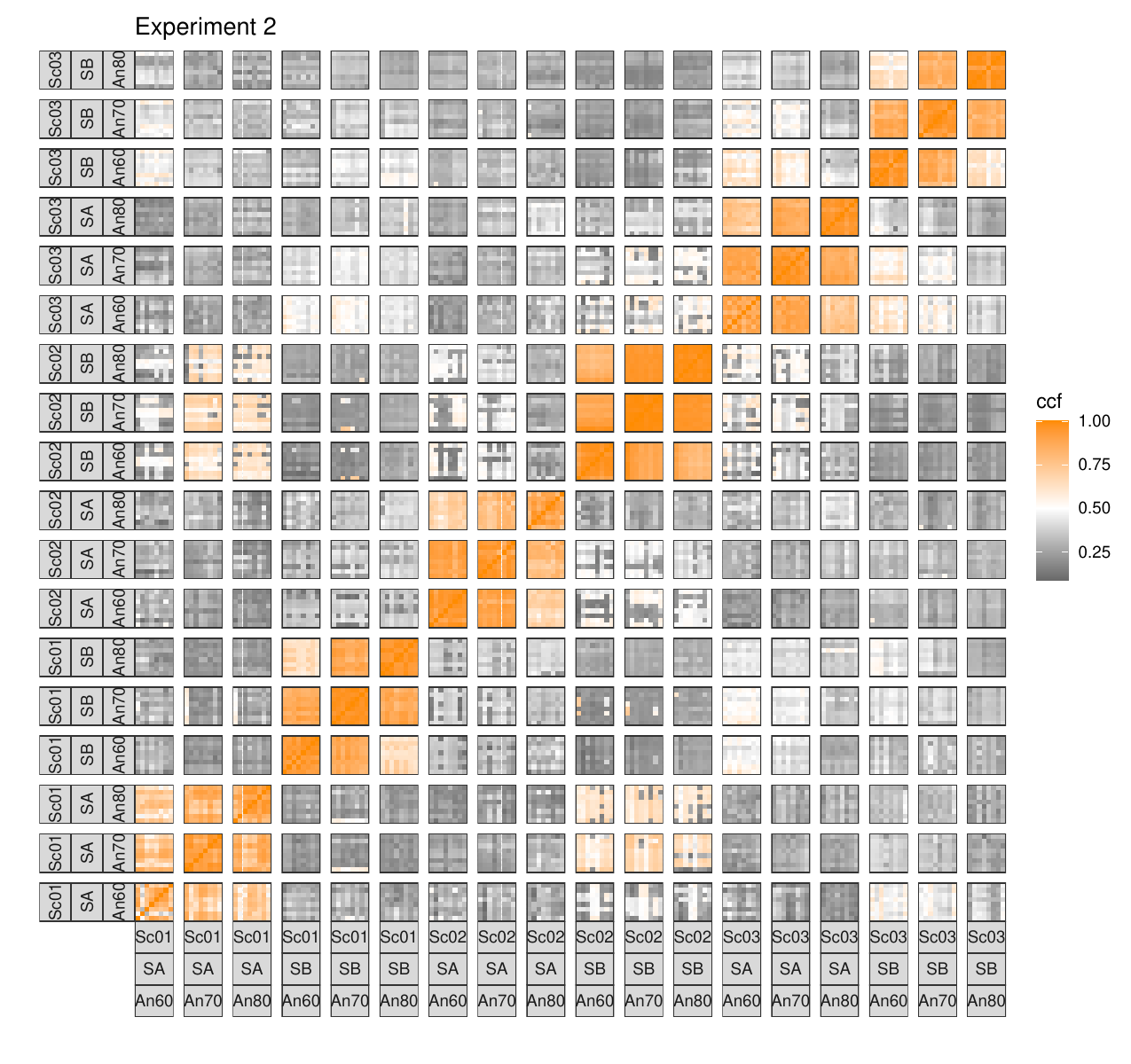} 

}

\caption[Similarity matrix displayed as a heat map for the similarities between toolmarks made by 3 large screwdrivers, sides A and B, at 3 different angles (60, 70, 80), with 8 replicates each]{Similarity matrix displayed as a heat map for the similarities between toolmarks made by 3 large screwdrivers, sides A and B, at 3 different angles (60, 70, 80), with 8 replicates each.}\label{fig:simmatrix_exp2}
\end{figure}

\end{knitrout}


\begin{knitrout}
\definecolor{shadecolor}{rgb}{0.969, 0.969, 0.969}\color{fgcolor}\begin{figure}

{\centering \includegraphics[width=\textwidth]{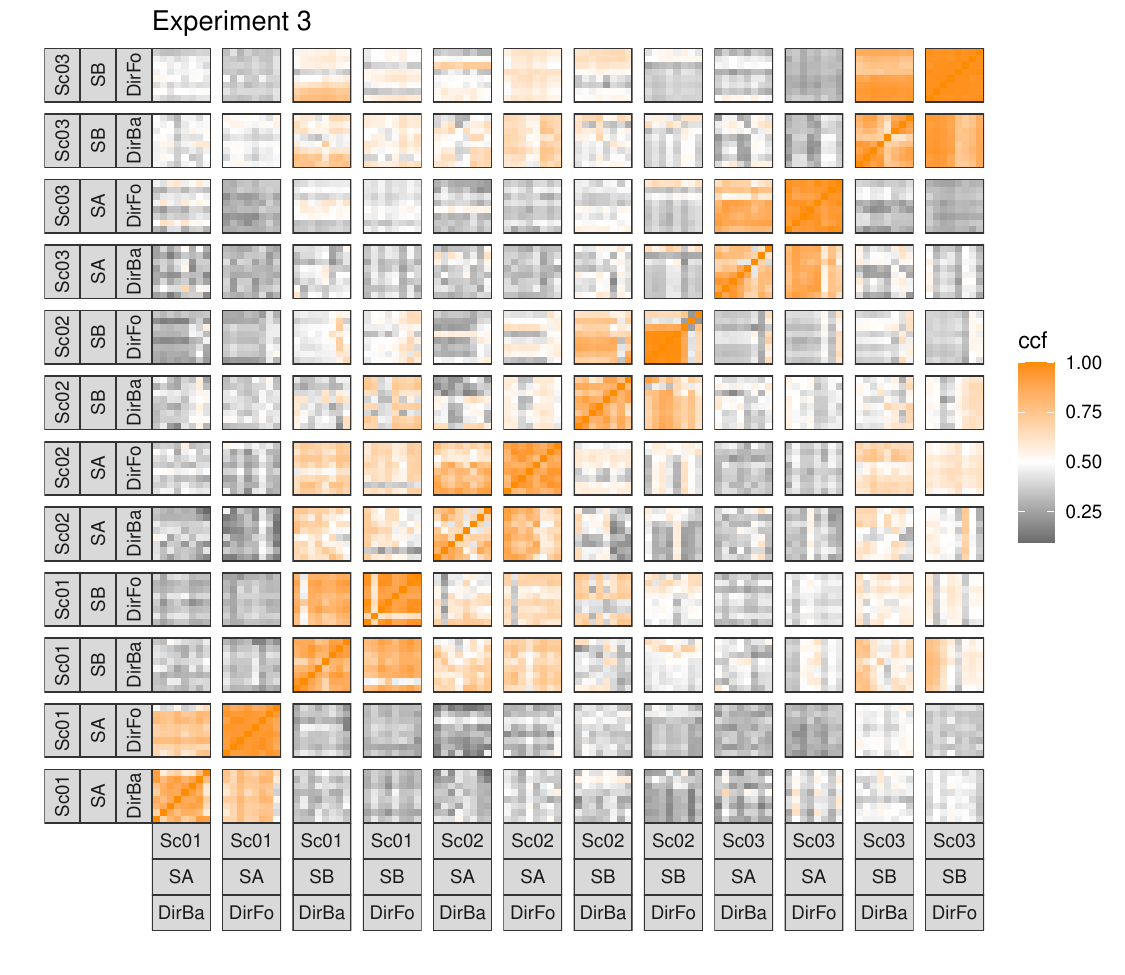} 

}

\caption[Similarity matrix displayed as a heat map for the similarities between toolmarks made by 3 small screwdrivers, sides A and B, at two different directions (pushing and pulling), with 8 replicates each]{Similarity matrix displayed as a heat map for the similarities between toolmarks made by 3 small screwdrivers, sides A and B, at two different directions (pushing and pulling), with 8 replicates each.}\label{fig:simmatrix_exp3}
\end{figure}

\end{knitrout}

\begin{table}[h] 
\centering
\caption{Clustering results from PAM algorithm.}
\begin{tabular}{cccc} 
\toprule
 & Experiment 1 & Experiment 2 & Experiment 3 \\
\midrule
Clusters & 49 &  6  & 6 \\
\bottomrule
\end{tabular}
\label{tab:clusteringresults}
\end{table}

As shown in Table~\ref{tab:clusteringresults}, for experiment 1, the PAM algorithm finds 49 clusters. The expected number was 40, since experiment 1 has signals made by 20 different tools, by sides A and B, at the same angle and direction. Here, the Silhouette score reached a plateau after $k = 40$, which suggests that after 40 clusters the differences between groups are similar to the differences within groups. For experiments 2 and 3, the PAM algorithm finds 6 clusters each. Although there were 3 tools, with 2 sides each (for a total of 6 sources), we did not know whether the signals at different angles and directions would cluster together, or would be their own clusters. Indeed, the signals cluster by tool-side, not by angle or direction. Although it does not solve the ``degrees of freedom'' problem completely because there could be different clustering results at other settings, this is encouraging because it means that even at different angles and directions, these factors do not affect the clustering by source.

\subsection{Method 2: Densities}

One issue with generating the densities of KM and KNM is that our data has dependencies that arise due to replicate marks being generated by the same source. Ignoring this fact could lead to a biased density for the KNM scores. Others have tried plotting a naive pair of densities that assumes all pairs are independent and downsampling the KNM density so it has the same sample size as the KM density \citep{ommen, veneri2023ensemble}. We address this by averaging the correlations by source across replicates since this removes the dependencies by providing a single number per source. We also tried the two other methods mentioned, as shown in the Appendix, but we believe that averaging by source gives a more honest and careful approach to dealing with the dependencies in the data.

\begin{knitrout}
\definecolor{shadecolor}{rgb}{0.969, 0.969, 0.969}\color{fgcolor}\begin{figure}[H]

{\centering \includegraphics[width=.8\textwidth]{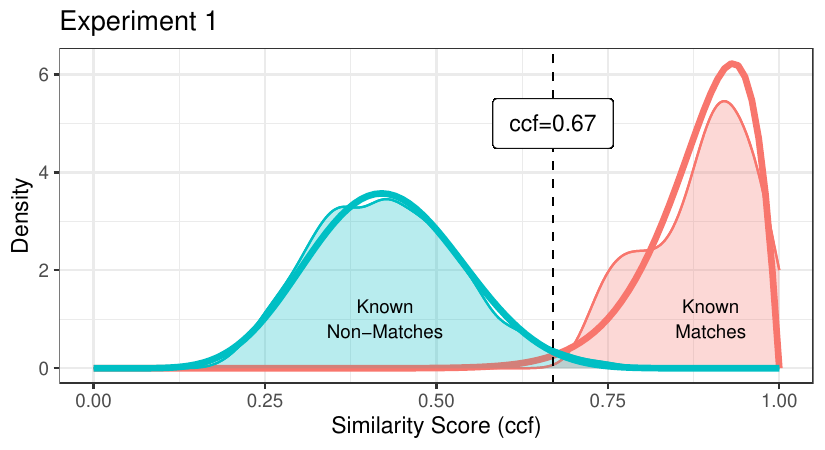} 

}

\caption[Densities for Known Match (KM) and Known Non-Match (KNM) pairs in terms of similarity (cross-correlation function, normalized to range from 0 to 1) between pairs]{Densities for Known Match (KM) and Known Non-Match (KNM) pairs in terms of similarity (cross-correlation function, normalized to range from 0 to 1) between pairs. The thick curves are Beta distributions that cross at the dashed line.}\label{fig:densities-exp1}
\end{figure}

\end{knitrout}

\begin{knitrout}
\definecolor{shadecolor}{rgb}{0.969, 0.969, 0.969}\color{fgcolor}\begin{figure}[H]

{\centering \includegraphics[width=.8\textwidth]{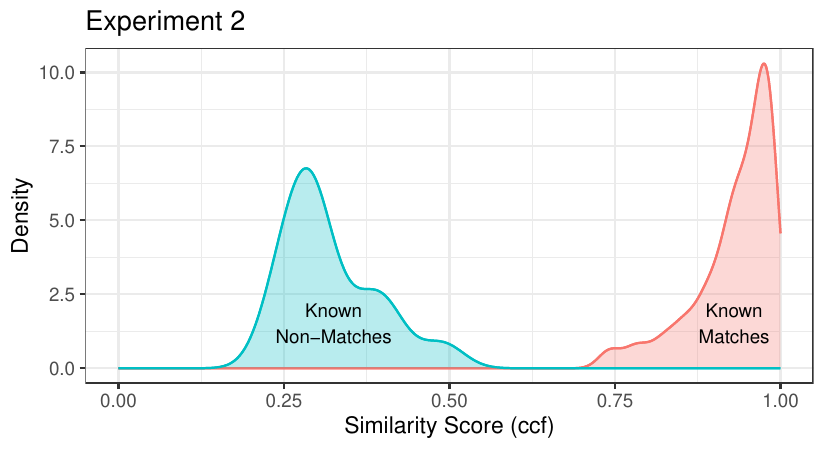} 

}

\caption[Densities for Known Match (KM) and Known Non-Match (KNM) pairs for Experiment 2]{Densities for Known Match (KM) and Known Non-Match (KNM) pairs for Experiment 2.}\label{fig:densities-exp2}
\end{figure}

\end{knitrout}

\begin{knitrout}
\definecolor{shadecolor}{rgb}{0.969, 0.969, 0.969}\color{fgcolor}\begin{figure}[H]

{\centering \includegraphics[width=.8\textwidth]{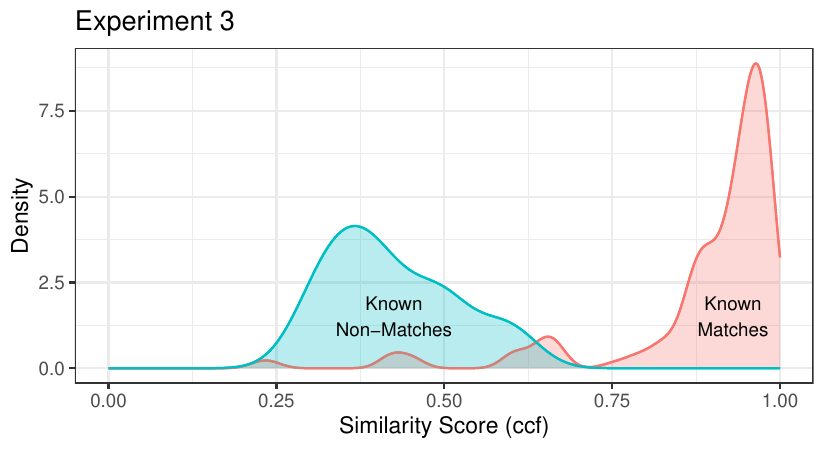} 

}

\caption[Densities for Known Match (KM) and Known Non-Match (KNM) pairs for Experiment 3]{Densities for Known Match (KM) and Known Non-Match (KNM) pairs for Experiment 3.}\label{fig:densities-exp3}
\end{figure}

\end{knitrout}

Figure~\ref{fig:densities-exp1} shows the densities of similarity scores for known matches (KM) and known non-matches (KNM) for experiment 1. Note that the similarity score is the cross-correlation function (ccf), normalized to range between zero and one by adding one and dividing by two. For the KNM density, we group similarity scores by their sources and average across replicates. In other words, for the KNM density, we consider all possible combinations of 8 replicates for one source across 8 replicates for the other, resulting in 64 similarity scores for a pair of sources, and then we take the average over these 64 scores. For the KM density, we include all pairwise combinations of replicates from each source. The Beta distributions are used to estimate likelihood ratios (see Section \ref{sec:likelihoodratio}). In Figures~\ref{fig:densities-exp2} and \ref{fig:densities-exp3}, we plot the same for experiments 2 and 3. However, we do not include the distributions or the threshold for these because they are based on a small dataset and we do not recommend classifying based on either of these experiments alone.

In experiment 1 (Figure~\ref{fig:densities-exp1}), the densities are well separated. The threshold, where the KNM and KM densities intersect is at 0.67. This can be used as a threshold for classification, i.e., above ccf=0.67 the pair is classified as same-source, and below as different-source. This result shows that toolmarks generated by consecutively manufactured screwdrivers really are more similar to each other if made by the same source and quite different from each other if made by different sources. 

In experiments 2 and 3 (Figures~\ref{fig:densities-exp2} and \ref{fig:densities-exp3}), we see that the densities are also separated, more clearly for experiment 2 than 3. These plots were made with much smaller datasets, so they are likely to be less smooth. Our interpretation of this separation is that the differences in angle and direction of mark generation are smaller than the differences between sources. Toolmarks made by the same source are more similar to each other than marks made by different sources, regardless of the angle and direction. This implies that the ``degrees of freedom'' problem is not actually so grave for toolmarks (within the range of 60-80 degrees) because marks made by the same source really are more similar to each other than marks made by different sources, despite the angle and direction of the tool.  According to the literature \citep{baiker2015}, the influence of angle is much larger for example in the range 80 to 110 (over the top) or 30 to 10 (very flat). Note that in Figure \ref{fig:simmatrix_exp2}, one can observe a decrease in the correlation score when 60 is compared to 80. This trend may continue when higher angle differences are introduced. Future research could explore this trend.

\subsection{Method 3: Likelihood ratio}\label{sec:likelihoodratio}

For experiment 1, we fit parametric (Beta) distributions on each curve. Selecting Beta distributions has many advantages \citep{song2018estimating}. For instance, the distribution ranges from 0 to 1, which is convenient for modeling probabilities. In addition, it has two parameters, $\alpha$ and $\beta$, which allow the distribution to take on many different shapes, some of which look similar to our KM and KNM densities. The parameters for Beta distributions are chosen such that the first moment and the second moment are matched with data. Figure~\ref{fig:densities-exp1} demonstrates the two density curves along with their fitted Beta distributions. Note that the parameters for Beta distributions are derived to match the first and second moment of the data. Specifically, we obtained that for the KM curve, $\alpha = 15.7494, \beta = 2.0665$ and for the KNM curve, $\alpha = 8.5774, \beta = 11.4628$. The curves are well separated, and they intersect around a similarity score (ccf) of 0.67.


\begin{figure}[h] 
   \centering
   \includegraphics[width=.6\textwidth]{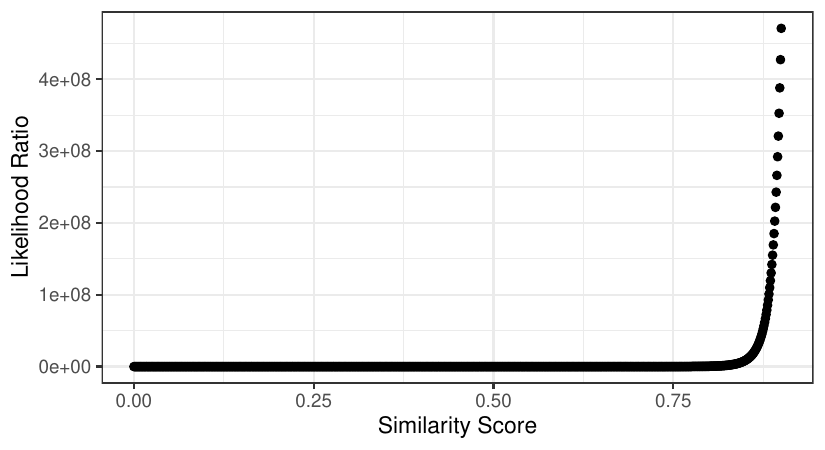} 
   \caption{Likelihood ratio as a function of the similarity score, for experiment 1. The likelihood ratio are calculated based on the fitted Beta distribution.}
   \label{fig:likelihoodratio}
\end{figure}


Figure \ref{fig:likelihoodratio} shows the likelihood ratio as a function of the Pearson correlation score between pairs of signals. Note that the likelihood ratio is quite small when the correlation is below 0.67, but it quickly surges once the correlation score is above 0.8. This gives us a statistical way to validate whether the given two toolmarks are indeed from the same source or not, based on the training data. As a result, for any new pair of toolmark signals, we can calculate a LR that will indicate whether the evidence is more likely under the ``same source'' proposition or the ``different source'' proposition.

We do not fit parametric distributions to the plots made with experiment 2 and 3 data, or classify based on their thresholds, because the densities are fit with small amounts of data, and we believe that the parametric assumptions based on these would be too strong. Such small datasets should not be used for statistical analysis.

\section{Method performance}\label{ref:performance}

\subsection{Cross-validated classification performance}

How well does classification work if we use the intersection point for the KM and KNM densities from experiment 1 as a threshold to classify between different-source and same-source pairs? Using the data from experiment 1, we use cross-validation to evaluate the classification performance. Specifically, we do the following:
(1) Split the data into two folds, one with all the marks made by the tools marked by even numbers and the other with all the marks made by tools with odd numbers.
(2) Calculate the threshold using half the data (we call this the ``training'' set).
(3) Use this threshold to classify between same-source and different-source pairs in the remaining data (the ``testing'' set.
(4) Calculate the sensitivity and specificity of our algorithm by comparing the predicted outcome with the ground truth (KM or KNM).
(5) Repeat this procedure with the second fold.
(6) Average the sensitivity and specificity to obtain the cross-validated performance metrics.

We repeat this procedure by only using the data from experiment 2 and only using the data from experiment 3, with their respective thresholds. We did this so we could test whether the different angles and directions lead to having such different marks that they are classified as different sources. 

\begin{table}[h] 
\centering
\caption{Cross-validated classification performance.}
\begin{tabular}{cccc} 
\toprule
            & Experiment 1 & Experiment 2 & Experiment 3 \\
\midrule
Sensitivity & 0.98 &  0.93  & 0.85   \\
Specificity & 0.96 &  0.98  & 0.95    \\
\bottomrule
\end{tabular}
\label{tab:performance}
\end{table}

Table \ref{tab:performance} shows the cross-validated sensitivity and specificity of our classification procedure. The table shows that the in-sample performance when using the threshold from experiment 1 (0.67) is very high. When using only the data from 3, the sensitivity is lower. It is difficult to say whether the lower sensitivity is due to having less data. However, after qualitatively observing the orange squares in the off-diagonals in Figures~\ref{fig:simmatrix_exp2} and \ref{fig:simmatrix_exp3}, it is not surprising to see that the classification is lower. Nevertheless, the performances are generally high for all three experiments.

\begin{figure}[h]
   \centering
   \includegraphics[width=.5\textwidth]{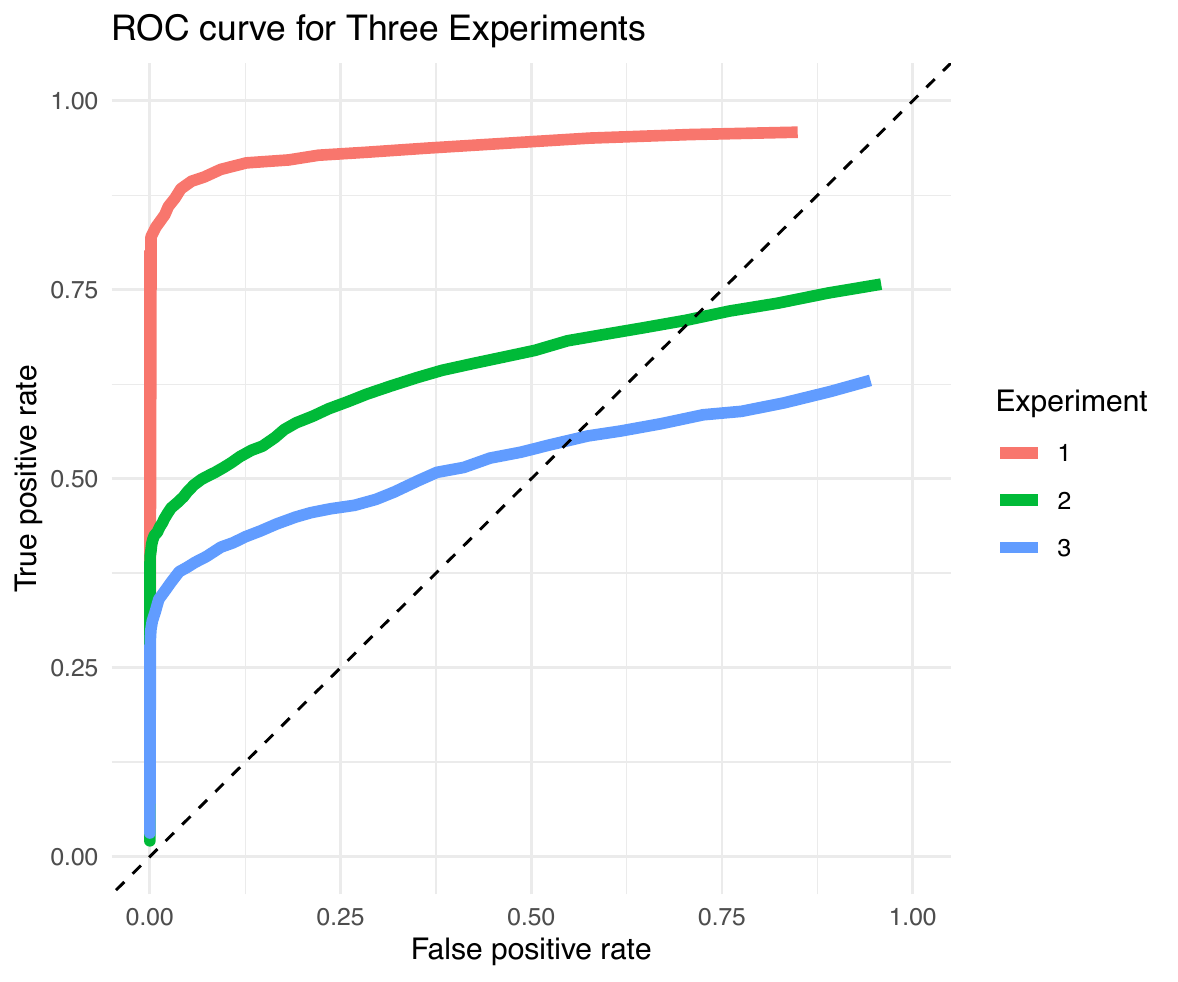} 
   \caption{Receiver operating characteristic (ROC) curve for the classifiation method based one the threshold of the KM and KNM density intersection for the three experiments.}
   \label{fig:roc}
\end{figure}

Figure \ref{fig:roc} shows the receiver operating characteristic (ROC) curve, which illustrates the performance of the binary classifier model at varying threshold values for the intersection of the KM and KNM curves. The experiment 1 threshold method is better at classifying, which makes sense since it is based on more data. Again, we do not recommend using experiments 2 and 3 to classify because they are based on too few data points.

\subsection{Performance as a function of length}

Sometimes practitioners have a very short striation mark from a tool, from a crime scene, and they need to know whether a candidate tool created this short mark. For example, the examiner may need to compare a short mark made by a slotted screwdriver on the edge of a surface, to longer test marks made by a candidate screwdriver. Another example is a thin wire that is cut with wire cutters -- the mark on the wire can be quite short. We can test the performance of the method as the length of one of the marks decreases.

Intuitively, as the length of one of the signals decreases, there is less information in the signal. Thus, the signal becomes more similar to the marks made by other tools, and the false positive rate increases. In other words, in the extreme case, the signal is very short. Thus, it has very little information, and this means it could have been produced by any of the candidate tools.

Figures~\ref{fig:performance-length-plots3} show the specificity and sensitivity as functions of length. That is, the performance of the classification method (using the data from Experiment 1, since it had the largest training dataset) as we shorten the length of one signal and compare it to a full-length signal. The sensitivity is not very informative -- there seem to be no major changes in the true positive or false negative rates as signal length decreases. 

A specificity of 0.9, which corresponds to a 10\% false positive rate, is reached at a signal length of about 1.5 mm. Once the signal length decreases further, there is not enough information to determine its source with any reasonable accuracy. At a signal length of just under 1 mm specificity drops to 50\%, i.e. the accuracy of assessing whether a signal of that length matches a specific source is equal to a coin flip. Note that reducing the signal length could have caused FP errors in the location of the shorter segment due to alignment (i.e., registration). Studying the location of where the shorter segments found high CCF would be a useful exercise. 

There is a similar dependence of accuracy on signal length in \cite{hare2017algorithmic}. The authors find that at 37.5\% of a land engraved area (corresponding to about $.375 \times 2.2$ mm = 0.825 mm) specificity drops below 0.9 (for a sensitivity of about 0.7).


\begin{knitrout}
\definecolor{shadecolor}{rgb}{0.969, 0.969, 0.969}\color{fgcolor}\begin{figure}[H]

{\centering \includegraphics[width=.8\textwidth]{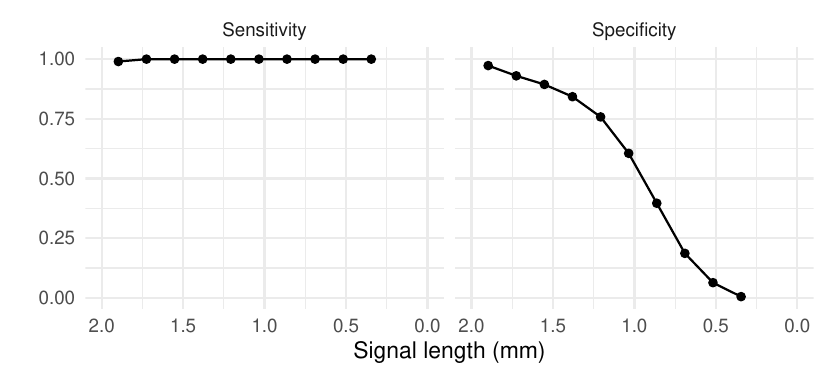} 

}

\caption[Sensitivity and specificity as a function of length]{Sensitivity and specificity as a function of length.}\label{fig:performance-length-plots3}
\end{figure}

\end{knitrout}


\section{Discussion}\label{sec:discussion}

Our method shows that it is possible to distinguish between same-source and different-source pairs of toolmarks reliably. We present an objective method to perform forensic toolmark comparisons that presents results using likelihood ratios and addresses the problem of the ``degrees of freedom''. We find that the changing the angle of attack (from 60 to 70 to 80 degrees with respect to the surface) and direction of mark generation (pulling and pushing) does not affect whether signals made by a source cluster together, or whether the algorithm classifies reliably. Our method has cross-validated sensitivity of 98\% and specificity of 96\% (note that this is for experiment 1). This is encouraging because it shows that the method performs well even with consecutively manufactured tools, which are products of the same manufacturing process and thus one of the most difficult groups to classify correctly. 

Our method holds for slotted screwdrivers and for screwdrivers that are made with a similar production method. To use this method to classify different types of tools, it may be necessary to perform similar experiments with different types of tools. A large database of 3D toolmarks could be created, and the more tools that researchers add, the better the method will be at classifying same- and different-source marks. Future research is needed to determine where the limits are of classifying using our training data.

We found that very short signals, below 1.5 mm in length, cannot be compared reliably, even in this experimental setup where we have high-quality 3D data generated under controlled conditions. These results are particularly relevant for establishing error rates for comparisons with respect to the length of a signal.

Since our screwdrivers were consecutively manufactured, any sub-class characteristics generated by the manufacturing process, i.e., marks that were common among screwdrivers despite the fact that they are different tools, are likely included in our data. 

In future work, it would be useful to study how generating toolmarks at smaller angles of attack (i.e., less than 60) affects the classification performance, as well as other analyses that we presented with data from experiment 2. \cite{baiker2014} had shown that smaller differences in angle did not affect classification, and our results agree with this, but that larger differences did affect classification. It would also be interesting to study the difference between making toolmaks with large and small screwdrivers. In our experimental setup, our mechanical rig only allowed us to make toolmarks at different angles with large screwdrivers, and toolmarks in different directions with small screwdrivers. Thus, it would be useful to have a complete factorial design, where toolmarks are generated wiht a different mechanical rig. Furthermore, it would be interesting to include degree of force applied as one of the factors, and how the factors interact with each other. We do not know how factors, such as angle of rotation, interaction of one tool with another as in wire cutters, tools that have strong class characteristics like serrated knives affect the classification performance of the algorithm out-of-sample. This study was conducted using lead, as lead was the material we found to capture marks of the entire screwdriver tip. This is a a foundational study. It would be interesting to study how the marks change with different substrate materials, in addition to simply exhibiting marks of the incomplete screwdriver tip. It would also be interesting to study the effects on classification due to degradation of the tool with use and time.

Although commonly used in forensics, classification by using a threshold from the density plots can be not a great idea because the threshold relies on parametric assumptions, it gives great importance to the tails of the distributions where there is the least amount of data available, and it is not clear how much the model performance depends on the training data. Would it help to diversify the training data for classification? And in what way? Would it be better to include toolmarks from a variety of different tools in the same training data, in a ``kitchen sink'' style? We did include the data from all three experiments in training densities, but we did not include them in this article because since experiments 2 and 3 have so much less data than experiment 1, the results did not change much from those for experiment 1. These questions remain for future research. It would be worthwhile to study the generalizability of this threshold to other tools and factors. Some methods have been trained on specific screwdrivers, there is evidence that they could be used to compare other tools as well \citep{baiker2014}. This is promising because it means that, for toolmark comparisons to be accurate and useful, it might only be necessary to train algorithms using a small number of tools. Collecting data from all the tools that could be used to commit a crime is practically impossible, especially as the number and types of tools grows over time. Further research is necessary to determine how much each type of tool generalizes to other types of tools.

\section{Conclusion}\label{sec:conclusion}

In response to the 2009 NAS report and the 2016 PCAST report, we propose an objective method to perform toolmark comparisons. This method produces probabilistic results, it is consistent in its performance, and it is transparent. Furthermore, we have evidence that this algorithm is robust to changes in two factors, angle (at least to 20 degrees difference) and direction (pushing and pulling). We generated three original datasets of 3D toolmarks and their corresponding 2D signals, which is available for other researchers to use (forthcoming). 

To compare two striation toolmarks with our classification method, one can obtain a 3D image of the toolmarks with any equipment that provides sufficient resolution (around one micrometer), extract the signals (see guidance in our data section \ref{sec:data}), and calculate their cross-correlation function (normalized to range between zero and one). Then, if the ccf is higher than 0.67, our method concludes that the pair was made by the same source, if it is lower than 0.67 it was made by two different sources. To obtain a probabilistic result, a likelihood ratio can be obtained using our fitted Beta distributions. There remains a question about how much these results (i.e., the threshold and the performance metrics) generalize to other factors and other types of tools.

The shift from subjective to objective comparison methods in pattern-matching forensic disciplines has the potential to improve consistency, allow for the demonstration of process validation, and allow for more transparency and more possibilities for validation. All of these can reduce errors in comparisons and, therefore, improve the criminal justice system.

\section{Acknowledgments - removed for blind review}


\bibliographystyle{plainnat}

\bibliography{toolmarks-second-submission.bib}

\newpage

\section{Appendix A}\label{sec:appendixa}

\begin{figure}[h]
\hfill
\subfigure[Title A]{\includegraphics[width=.45\textwidth]{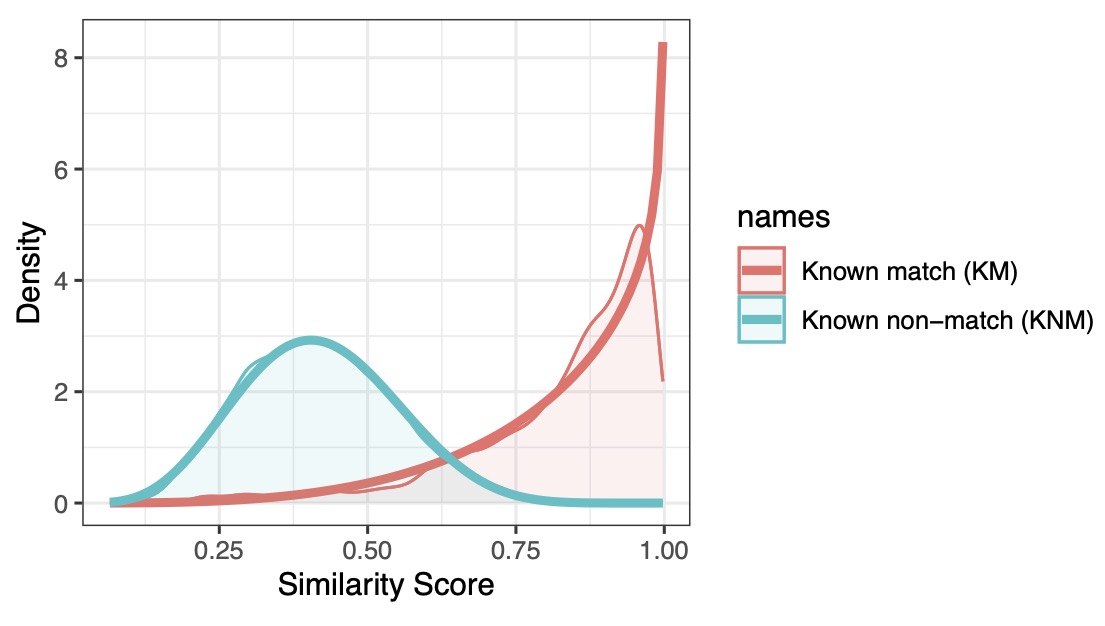}}
\hfill
\subfigure[Title B]{\includegraphics[width=.45\textwidth]{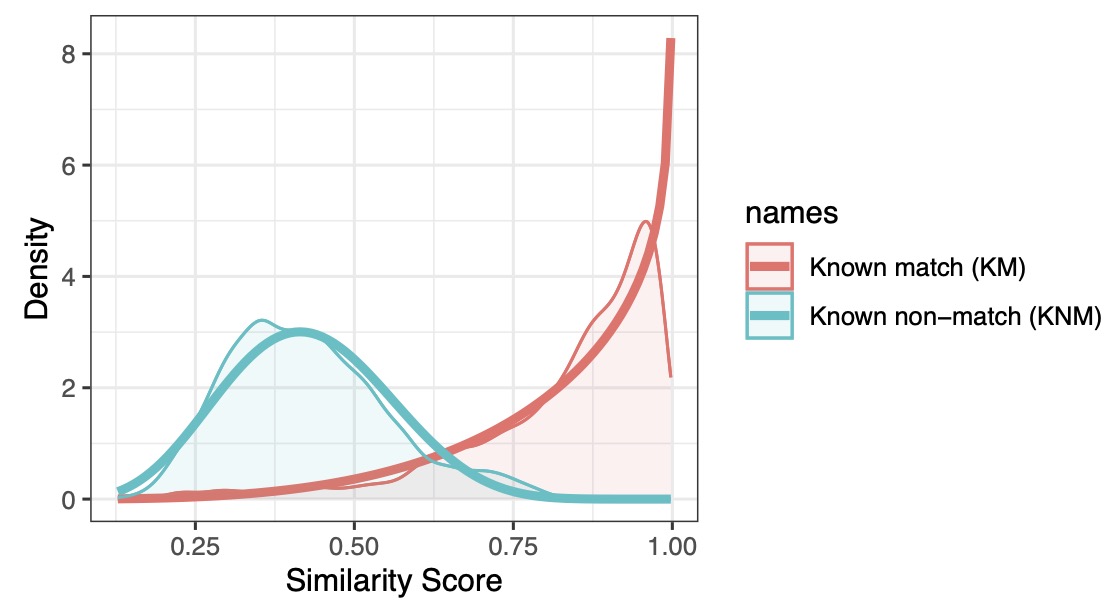}}
\hfill
\caption{Alternative approaches to deal with dependencies in the data.}
\label{fig:densities-v2-3}
\end{figure}

Figures \ref{fig:densities-v2-3}a and \ref{fig:densities-v2-3}b show the same densities as in Figure \ref{fig:densities-exp1}, but fit with a more naive approach. Figure \ref{fig:densities-v2-3}a includes all the KNM pairs, assuming they are independent and Figure \ref{fig:densities-v2-3}b only includes a random sample of points from the KNM pairs equal in sample size to the KM density. For both versions, the point at which the two densities cross is slightly lower than the version 1 value, 0.67, by about 5 percentage points at 0.64.

\end{document}